\pdfoutput=1

\documentclass[11pt]{article}

\usepackage[final]{acl}

\usepackage{tikz}

\usepackage{booktabs}

\usepackage{times}
\usepackage{latexsym}

\usepackage[T1]{fontenc}

\usepackage[utf8]{inputenc}
\usepackage{cleveref}

\usepackage{enumitem}
\setlist{noitemsep}

\usepackage{microtype}

\usepackage{inconsolata}

\usepackage{graphicx}
\usepackage[normalem]{ulem}

\usepackage{colortbl}

\newif\ifomitdeltext
\omitdeltexttrue

\definecolor{blunavy}{HTML}{0072b2}

\ifomitdeltext
  \newcommand{\SBdel}[1]{}
\else
  \newcommand{\SBdel}[1]{{\color{blunavy}\sout{#1}}}
\fi

\usepackage{enumitem}
\usepackage{tcolorbox}
\usepackage{subcaption}
\usepackage{adjustbox}

\title{When LLMs Can't Help: Real-World Evaluation of LLMs in Nutrition}

\author{
 \textbf{Karen Jia-Hui Li\textsuperscript{1,2}}\quad
 \textbf{Simone Balloccu\textsuperscript{3}}\quad
 \textbf{Ondrej Dusek\textsuperscript{1}}\quad
 \textbf{Ehud Reiter\textsuperscript{4}}
\\
 \textsuperscript{1}Charles University, Faculty of Mathematics and Physics, Czechia\\
 \textsuperscript{2}Saarland University, Germany\qquad
 \textsuperscript{3}TU Darmstadt, Germany\\
 \textsuperscript{4}University of Aberdeen, Scotland, UK\\
 \small{
   \textbf{Correspondence:} \href{mailto:li.karen.jh@gmail.com}{li.karen.jh@gmail.com}
 }
}

\begin{document}
\maketitle
\begin{abstract}

The increasing trust in large language models (LLMs), especially in the form of chatbots, is often undermined by the lack of their extrinsic evaluation. This holds particularly true in nutrition, where randomised controlled trials (RCTs) are the gold standard, and experts demand them for evidence-based deployment. LLMs have shown promising results in this field, but these are limited to intrinsic setups. We address this gap by running the first RCT involving LLMs for nutrition. We augment a rule-based chatbot with two LLM-based features: (1) message rephrasing for conversational variety and engagement, and (2) nutritional counselling through a fine-tuned model. In our seven-week RCT (n=81), we compare chatbot variants with and without LLM integration. We measure effects on dietary outcome, emotional well-being, and engagement. Despite our LLM-based features performing well in intrinsic evaluation, we find that they did not yield consistent benefits in real-world deployment. These results highlight critical gaps between intrinsic evaluations and real-world impact, emphasising the need for interdisciplinary, human-centred approaches.\footnote{We provide all of our code and results at: \\ \href{https://github.com/saeshyra/diet-chatbot-trial}{https://github.com/saeshyra/diet-chatbot-trial}}
\end{abstract}

\section{Introduction}

Every day, individuals make over 200 food-related decisions \citep{Wansink2007,van2016food}. With sedentary lifestyles becoming increasingly common \citep{Park2020} and global health issues on the rise \citep{Malik2013, Rowley2017}, scalable interventions are needed. Digital health technologies via mobile devices offer accessible solutions \citep{Vearrier2018,Senbekov2020}. 

In parallel, advances in fine-tuned language models enabled the generation of human-like responses for many practical applications \citep{Wei2022, Min2024}. This resulted in a general hype and trust—especially among laypeople and companies—in the potential of this technology~\citep{strange2024three}. This also applies to nutrition: LLMs look promising for tasks like meal recommendation, providing dietary advice, and general domain understanding \citep{Niszczota2023, Naja2024, Tsiantis2024}. Randomised controlled trials (RCTs) are required by domain experts before any real-world deployment~\citep{stolberg2004randomized,Hariton2018,Baumel2019}, as they are the foundation of evidence-based medicine and give objective measures of real-world impact. However, past evaluations of LLMs in nutrition are intrinsic only. No evidence has been collected regarding the impact of LLMs in real-world nutrition tasks. This includes sustained diet coaching, where users receive feedback on improving their dietary habits~\citep{vrkatic2022nutritional}, or nutritional counselling, where tailored empathetic support helps users address more complex dietary issues~\citep{vasiloglou2019challenges}.

We conduct the first extrinsic evaluation of an LLM-enhanced chatbot for these two tasks.  We start from a rule-based chatbot capable of scanning users' food diaries to provide tailored insights. Then, we integrate two LLM-based features: (1) a rephrasing module and (2) a nutritional counselling model. The former varies the base templated responses to make communication more engaging, while the latter provides tailored support, comfort, and suggestions for users' specific dietary concerns. In a seven-week RCT with 81 participants, we compare three groups: a group using the full set of features (insights+rephrasing+counselling), an intermediate group (insights+rephrasing), and a base group using only the rule-based chatbot (insights only). We measure dietary outcomes, emotional well-being, and engagement. Ethics details are presented in \Cref{app:ethics}.

Based on our results, the ``promise'' of LLMs in nutrition falls short in the real world: the LLM-based features had little to no effect on any of the measures we consider. Our study provides critical insights into the effectiveness of LLMs in nutrition, the safe deployment of these models, and the interdisciplinary challenges of applying them to sensitive domains.

\section{Related Work}

Digital health interventions can improve accessibility, cost-effectiveness, and patient-centred care \citep{Greaves2013, Mitchell2019, Taj2019}. For example, telemedicine allows remote consultations \citep{Barbosa2021, Totten2022}, while wearable devices enable continuous health monitoring \citep{Izmailova2018, Natalucci2023}. Mobile health interventions can be highly effective in terms of user adherence \citep{Kamal2015, Muller2016, Hoeppner2017, Lee2018, Oyebode2020}.

In nutrition, chatbots have emerged as a promising tool to promote healthy eating habits, offering food intake tracking~\citep{Graf2015, Kerr2016}, educational content, and motivational messaging \citep{Fadhil2017, Casas2018, Maher2020}. Recent advancements in pre-trained language models can further expand their capabilities in this domain, but not without shortcomings. Recent LLMs can generate meal plans and dietary recommendations based on user needs, but their performance decreases in more complex cases~\citep{Niszczota2023, Naja2024, Tsiantis2024}. 

Beyond meal guidance and recommendations, AI chatbots are being increasingly explored for their potential to provide empathetic support towards behaviour change. Negative emotions can lead to poorer nutritional choices \citep{Devonport2019, Gonzalez2022} and tailored advice can mitigate this mental load \citep{Balloccu2022a, park2024daily}. Chatbots can be approachable, calming, and display adequate therapeutic skills~\cite{Zhang2020, Beilharz2021, Vowels2024}. This can enhance engagement in sensitive contexts like body image, eating disorders, and relationship counselling. Artificial empathy and personalised interactions from chatbots~\citep{Stephens2019,Rahmanti2022} can help users in pursuing healthier habits. The therapeutic promise of chatbots also raises important ethical questions. There is still need for deeper integration of empathy, mental health sensitivity, and patient safety into chatbot design \citep{Stein2017,Anisha2024}. When evaluated by experts, AI nutritional support is often scientifically sound but potentially outdated,  inaccurate~\citep{Kirk2023}, or even harmful for more complex cases~\citep{balloccu-etal-2024-ask}.

\begin{figure*}[!htb]
    \centering
\includegraphics[width=\textwidth]{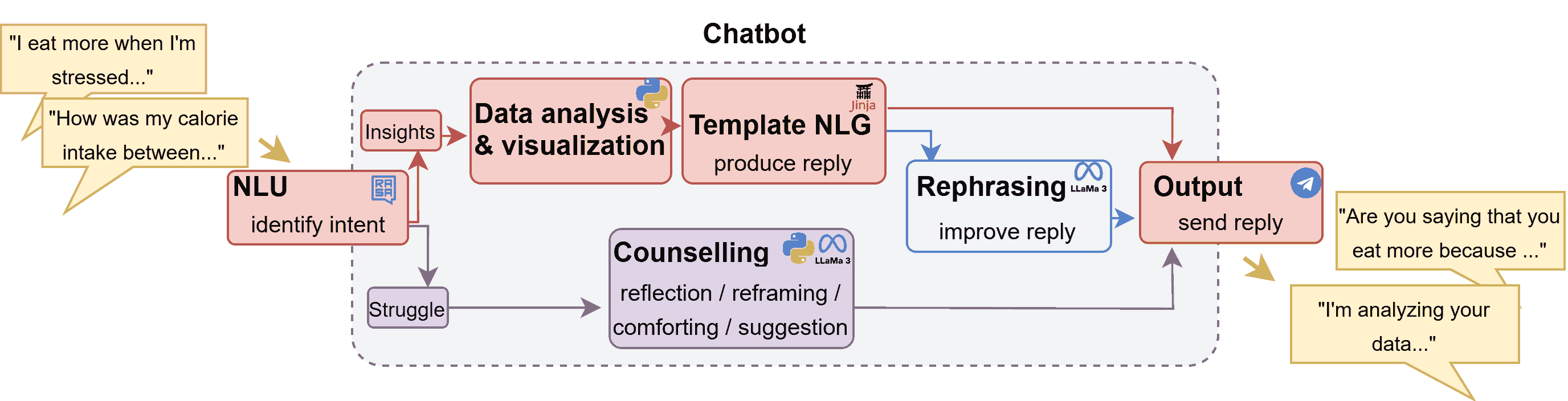}
    \caption{Overview of the chatbot architecture and functional flow. The \texttt{BASELINE} version uses the red flow only, \texttt{REPHRASED} adds the step marked in blue, and \texttt{FULL} adds the flow marked in purple. We provide an example of the insights flow in \Cref{fig:chatbot-outputs} and the supportive text flow in \Cref{tab:counselling-outputs}.}
    \label{fig:chatbot-arch}
\end{figure*}

While accuracy is important, user engagement also plays a critical role to the success of digital health interventions in nutrition. When users lose interest in using the chatbots, this causes a rapid decrease in dietary adherence, and eventually result in early drop-out~\citep{fadhil2018can, maher2020physical, balloccu2021unaddressed}. User satisfaction and motivation are usually pursued through personalisation, communication, visual elements~\citep{Kettle2024, Balloccu2022b}, gamification~\citep{Fadhil2017}, or social support mechanisms~\citep{Svetkey2015}.

For all of the above aspects, rigorous extrinsic evaluation is needed, to move chatbots from experimental prototypes to deployable healthcare assistants~\citep{Baumel2019}. Yet, existing evaluations are synthetic~\citep{mishra2024evaluation, yang2024chatdiet}, focused on textual characteristics or accuracy against standardised benchmarks~\citep{parameswaran2024optimizing,azimi2025evaluation}. Our work is, to our knowledge, the first~\citep{omar2024large} randomised controlled evaluation of LLM-delivered diet coaching and nutritional counselling over an extended deployment.

\SBdel{Current progress on more up-to-date related work:\begin{itemize}
    \item There seems to be no extrinsic evaluation on the use of LLMs in nutrition, neither wrt diet goal nor emotional affect (counselling etc). Besides not finding anything, \citep{omar2024large} offer a relatively recent (Nov 2024) overview of completed and ongoing clinical trial using LLMs in healthcare and none of them is about nutrition. 
    \item \citep{azimi2025evaluation} evaluated many recent LLMs on the registered dietitian exam ("LLMs pass the bar exam" style) showing overall "acceptable" results, depending on the tested topic, prompting strategy and model.
    \item Reading in progress: \citep{yang2024chatdiet,mishra2024evaluation,parameswaran2024optimizing,wang2023application,mishra2024evaluation} 
\end{itemize}}

\section{Chatbot Development}

\begin{figure}[ht!]
    \centering
    \begin{subfigure}[b]{0.75\linewidth}
        \centering
        \includegraphics[width=\linewidth]{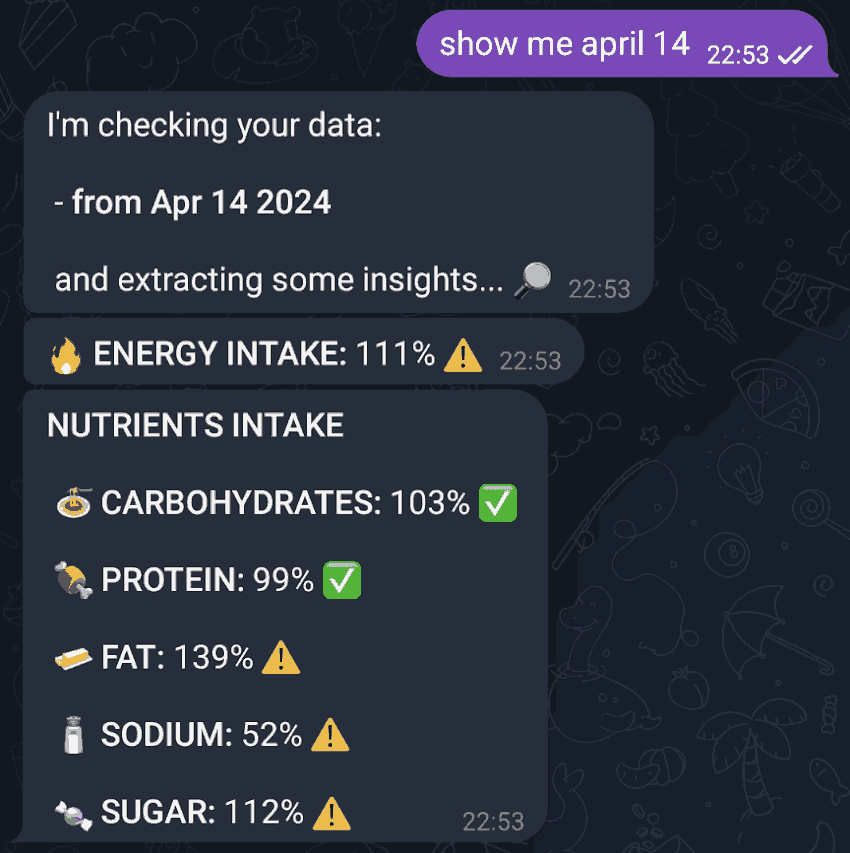}
        \caption{\textbf{Basic insights} into all monitored nutrients for a single day.}
        \label{subfig:basic-insights}
    \end{subfigure}
    \hfill
    \begin{subfigure}[b]{0.75\linewidth}
        \centering
        \includegraphics[width=\linewidth]{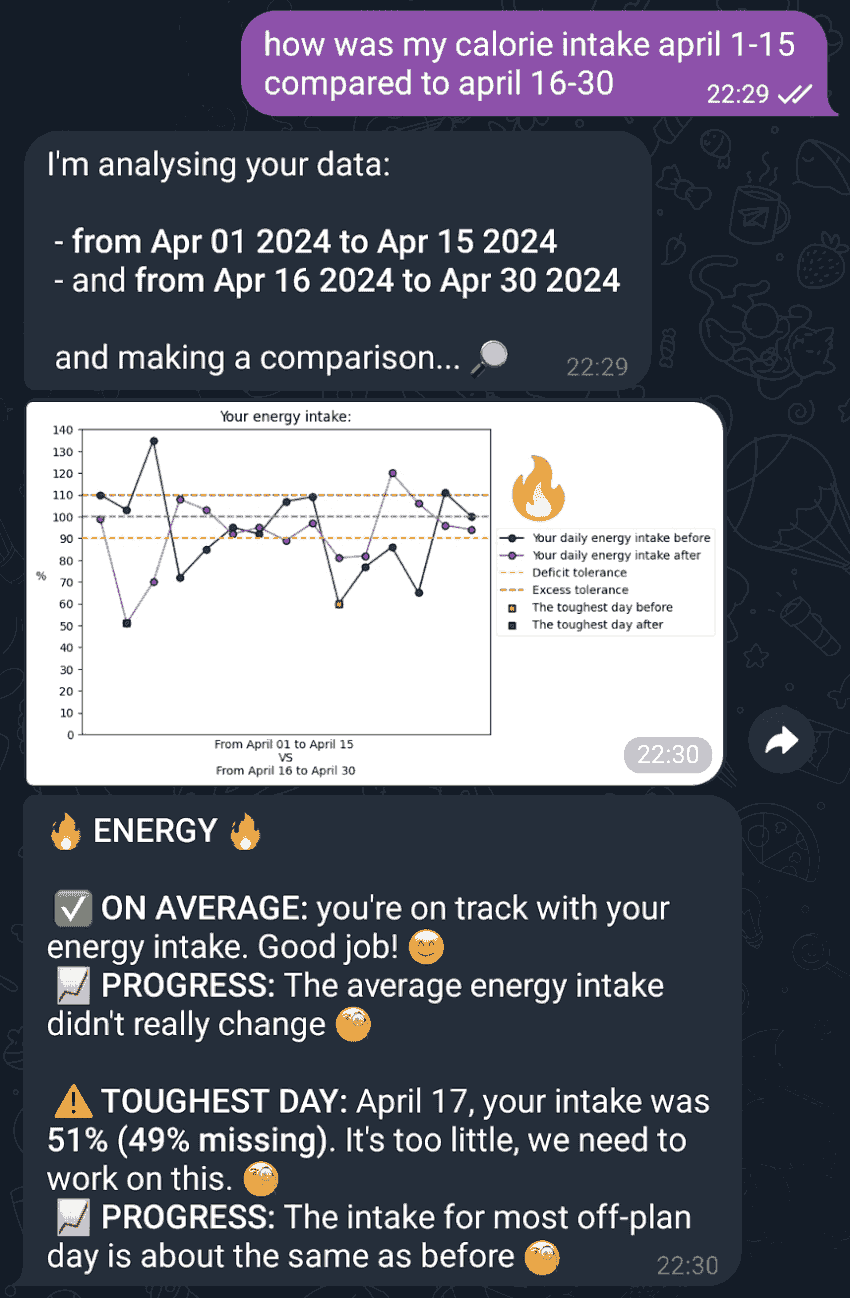}
        \caption{\textbf{Advanced insights}  comparing the average calorie intake and toughest day from the first half of the month to the other half.}
        \label{subfig:adv-insights}

    \end{subfigure}
    \caption{Examples of the chatbot outputs.}
    \label{fig:chatbot-outputs}
\end{figure}

We extend the rule-based diet-coaching chatbot by \citet{Balloccu2022b}, which is deployed on the Telegram messaging platform and designed to deliver personalised nutritional insights based on users' food diary on MyFitnessPal~\citep{evans2017myfitnesspal}, a popular and freely available calorie-counting app. The core functionality of the chatbot involves delivering insights in two forms: (1) \textit{basic insights} (\Cref{subfig:basic-insights}), which are simple recaps of a user's dietary intake—calories and nutrients, and (2) \textit{advanced insights} (\Cref{subfig:adv-insights}), which present an extended textual description with some corresponding data visualisation.

For this work, we extend the chatbot\footnote{We use \href{https://github.com/uccollab/philhumans-diet-coaching-chatbot}{the code} made available by the authors.} with two LLM-powered features: rephrasing to vary the templated responses, and nutritional counselling through fine-tuned models. \Cref{fig:chatbot-arch} illustrates the architecturenflow of our chatbot. Its main objective is to improve long-term diet adherence, emotional well-being, and user engagement.

\subsection{Rephrased Responses}

The original chatbot code included a small set of slots used to vary the templated responses. This system ensured consistency and safety, but lacked the conversational fluidity of human dialogue. To enhance communication variety, we prompt an LLM to rephrase the templated outputs, while maintaining clarity for more structured messages. These enhancements aim to enable more varied communication with the chatbot, and encourage higher engagement among trial participants.

To achieve high-quality rephrasing without the need for additional fine-tuning, we experimented with prompt engineering using instruction-tuned variants of Gemma 7B \citep{gemmateam2024}, Mistral 7B \citep{jiang2023mistral7b}, and Llama 3 8B \citep{llama3modelcard}, settling on Llama 3 8B as the production model.

\begin{figure}[t]
    \centering
    \begin{subfigure}[b]{1\linewidth}
        \centering
            \small



        \begin{tcolorbox}[colback=cyan!10!white, colframe=cyan!75!black, arc=3mm, boxrule=0.5pt, width=\textwidth, title= Final Rephrasing Prompt\\ \texttt{INTENT}: ``\texttt{compare\_no\_dates}'';
        \texttt{NUTRIENT}: \texttt{None};]
            The user requested comparative insights into their food diary but did not give dates. Do not greet the user. Do not include additional information. Use simple language and emojis. Rephrase the following message to the user.\\
            
            \textbf{Rephrase the message:}\\
            Please give me two dates or a date range to compare
        \end{tcolorbox}
        \label{fig:rephrasing-prompt-example}
    \end{subfigure}

    \begin{subfigure}[b]{1\linewidth}
        \small

        \begin{tcolorbox}[colback=cyan!10!white, colframe=cyan!75!black, arc=3mm, boxrule=0.5pt, width=\textwidth, title=Example Rephrased Output]
            To help you with your food diary, could you please provide two specific dates or a date range for comparison? \includegraphics[height=\fontcharht\font`\B]{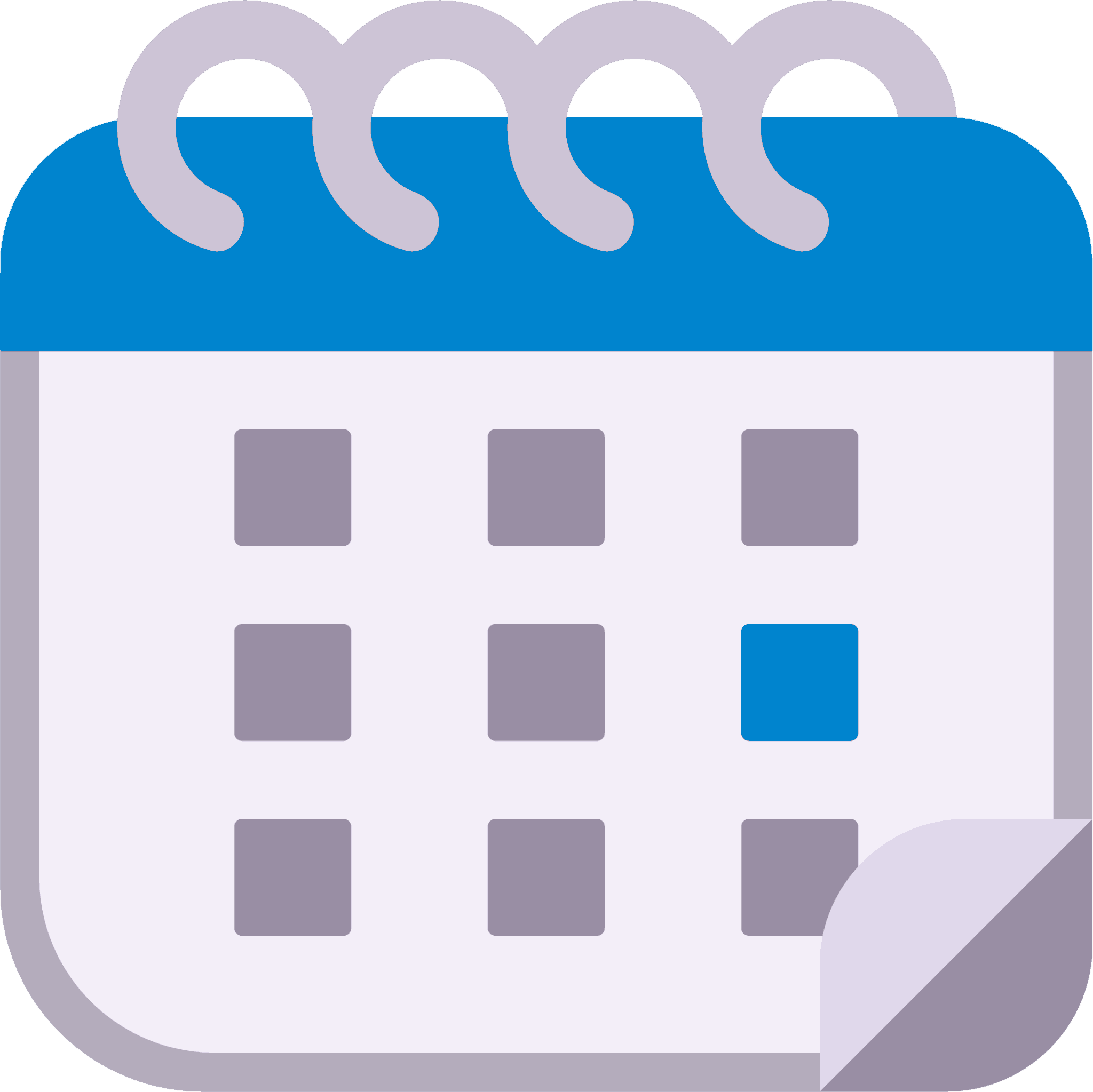}\includegraphics[height=\fontcharht\font`\B]{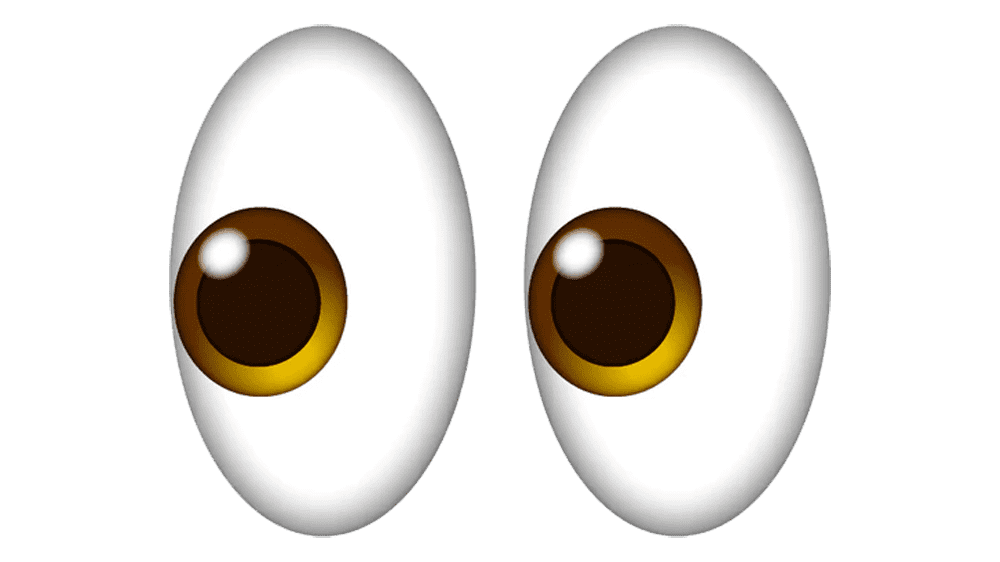}
        \end{tcolorbox}

        \label{fig:rephrased-output-example}
    \end{subfigure}
    \caption{An example of the dynamic rephrasing. The context leading up to the intent of the templated chatbot output (``\texttt{compare\_no\_dates}'' + no nutrient specified) is extracted from the NLU pipeline and dynamically added to the prompt, resulting in the rephrased output.}
    \label{fig:rephrased-example}
\end{figure}

With basic prompting (as in \Cref{fig:rephrasing-prompt-init}), we faced issues with ambiguous and context-sensitive message templates (\Cref{fig:rephrase-problem-output}). To address these, we exploit the fact that the rule-based chatbot gives us explicit access to the output message intent (e.g. insights on which nutrients, over which days). We develop a targeted prompt that dynamically adapts to message context with explicit, intent-specific instructions (\Cref{fig:rephrased-example}), reducing hallucinations by constraining the model to rephrase within context.

\begin{figure}[t]
    \small
    \centering
    \begin{subfigure}[b]{1\linewidth}
        \centering
        \begin{tcolorbox}[colback=cyan!10!white, colframe=cyan!75!black, arc=3mm, boxrule=0.5pt, width=\textwidth]
                \textbf{Rephrase the following message to the user}, keeping any mentioned dates. Do not introduce new dates or assume time periods. Do not add extra information. Use emojis.\\
                \textbf{\texttt{[message]}}
        \end{tcolorbox}
    \end{subfigure}
    \caption{Initial prompt for message rephrasing.}
    \label{fig:rephrasing-prompt-init}
\end{figure}

\begin{figure}[t]
    \centering
    \begin{subfigure}[b]{1\linewidth}
        \small
        \begin{tcolorbox}[colback=cyan!10!white, colframe=cyan!75!black, arc=3mm, boxrule=0.5pt, width=\textwidth, title=Example Templated Output]
            \includegraphics[height=1.2\fontcharht\font`\B]{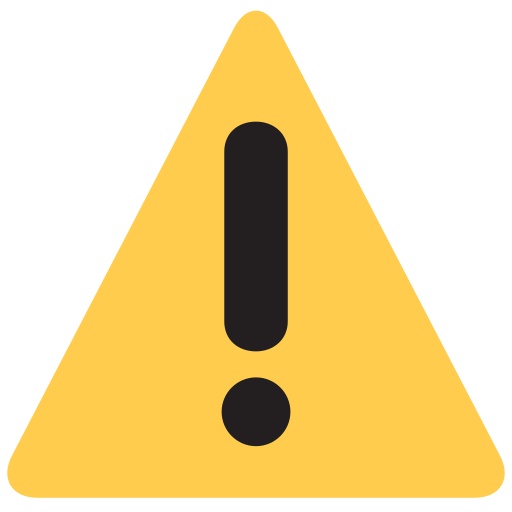} \textbf{TREND AND CONSISTENCY:} sorry, I need 3 or more days for this \includegraphics[height=1.2\fontcharht\font`\B]{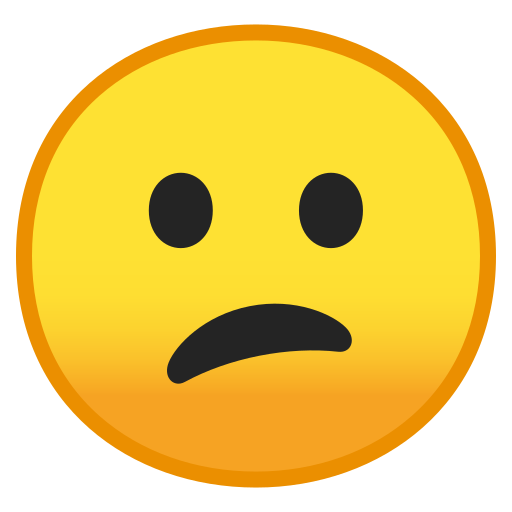}
        \end{tcolorbox}
        \begin{tcolorbox}[colback=cyan!10!white, colframe=cyan!75!black, arc=3mm, boxrule=0.5pt, width=\textwidth, title=Example Rephrased Output with Initial Prompt]
            \includegraphics[height=1.2\fontcharht\font`\B]{images/warning-emoji.png} Need a little time to get into the swing of things! \includegraphics[height=1.2\fontcharht\font`\B]{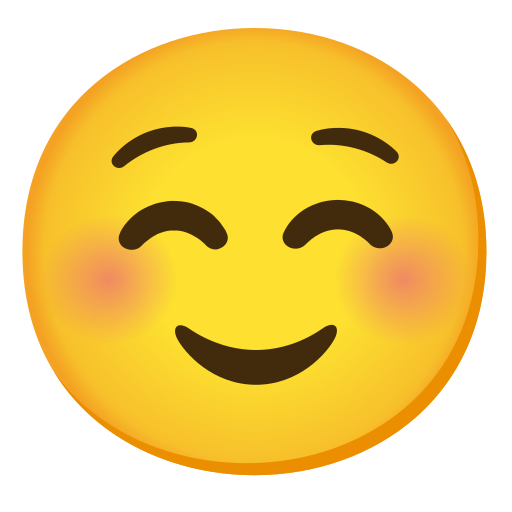} Please allow at least 3 days for this \includegraphics[height=1.2\fontcharht\font`\B]{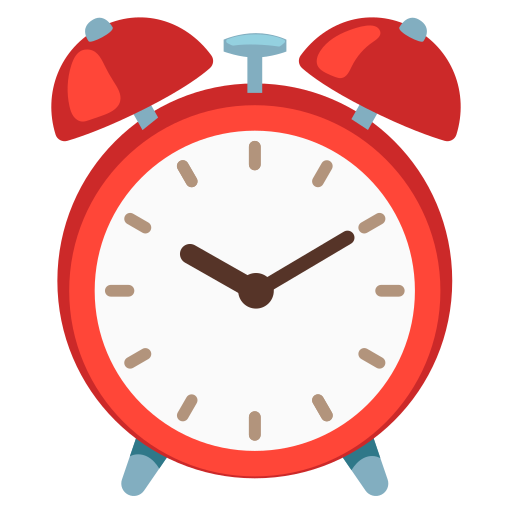}
        \end{tcolorbox}
    \end{subfigure}
    \caption{Example of a problematic rephrased output from the initial rephrasing prompt (\Cref{fig:rephrasing-prompt-init}), due to ambiguity in the original templated message responding to a user's request for advanced insights over a time period shorter than three days.}
    \label{fig:rephrase-problem-output}
\end{figure}

\begin{figure*}[!htb]
    \centering
    \includegraphics[trim=0.5cm 0.35cm 0.5cm 0.35cm, clip,scale=1.1]{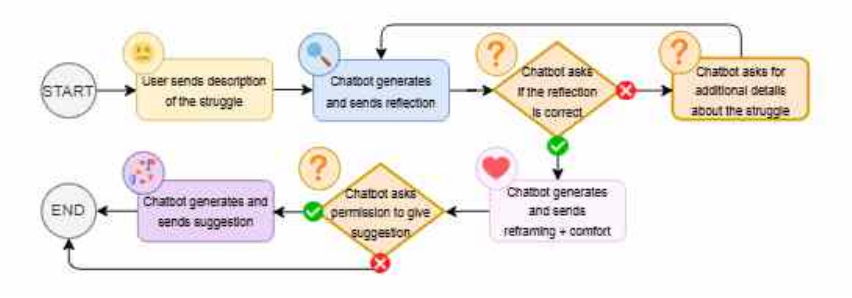}
    \caption{Nutritional counselling flow. When receiving a dietary struggle, the chatbot generates and sends reflections, asking the user for feedback (up to a limited amount of retries). Once a correct reflection is generated, comfort and reframing statements are generated and sent. Following this, the chatbot asks for permission to send a suggestion, ending the conversation.}
    \label{fig:counselling-flow}
\end{figure*}

\subsubsection{Evaluation}

We conducted an additional human evaluation with 20 native English-speaking crowd workers on Prolific. Participants were shown pairs of templated and rephrased messages and asked to indicate which they preferred, which felt more natural, and whether both conveyed the same meaning. About 65\% of responses were preferred in their rephrased form, and 72\% were judged more natural. Only a small number of participants (n=6) reported any differences in meaning between the messages. These findings confirm that LLM-based rephrasing, when carefully prompted, can enhance the linguistic quality and engagement of chatbot responses. We provide more information on this evaluation in~\Cref{app:rephrasing-eval}.

\subsection{Nutritional Counselling}
\label{ssec:nutr-couns}

We integrate a nutritional counselling feature designed to support users not only with dietary data, but also with the psychological and behavioural challenges of healthy eating. This feature is powered by a language model fine-tuned on the \texttt{HAI-Coaching} dataset \citep{balloccu-etal-2024-ask}, a collection of $\sim$2.4K crowd-sourced dietary \textit{struggles} paired with $\sim$100K expert-annotated supportive responses. The \textit{struggles} are textual descriptions of problems affecting people's diet and cover a wide variety of topics, from snacking and dietary restrictions, to emotional eating, anxiety, and depression. The responses are equally split into four categories—\textit{reflection} (understanding the struggle), \textit{comfort} (providing emotional support), \textit{reframing} (portraying the struggle positively), and \textit{suggestion} (actionable next steps)—as curated by human experts and reflective of the psychological research behind nutritional counselling.

We initially test by fine-tuning several models: GPT-2 medium \citep{gpt2}, FLAN-T5 base \citep{Chung2024}, BabyLlama \citep{Timiryasov2023}, Gemma 7B \citep{gemmateam2024}, Mistral 7B \citep{jiang2023mistral7b}, and Llama 3 8B \citep{llama3modelcard}. We include older and more limited models (GPT-2 and BabyLlama) to inspect whether newer, instruction-tuned ones offer a real advantage in terms of performance.

For GPT-2 and BabyLlama, we guide the generation of each category of supportive text via special tokens ("\texttt{<|struggle|>}, "\texttt{<|reflection|>}, etc.) in a controllable text generation fashion~\citep{keskar2019ctrl,li2021prefix,zhang2023survey}. For instruction-tuned models, we use category-specific prompts, mirroring those used to create the dataset~\citep{balloccu-etal-2024-ask}. We provide the prompts in \Cref{tab:finetuning-prompts}. 

Following common practices in NLG intrinsic evaluation~\citep{sai2022survey}, we calculate BLEU-1/3 and BLEURT (\Cref{tab:experimentation-results}), and also conduct a qualitative review of the generated outputs. We immediately excluded GPT-2, BabyLlama, Gemma, and Mistral, as they showed poor performance and consistently failed in producing useful supportive text. FLAN-T5 showed strong BLEU-1 performance, but frequently produced contradictory or irrelevant responses. Llama 3, on the other hand, delivered semantically coherent and contextually appropriate suggestions across a range of examples, achieving the best BLEU-3 and BLEURT scores. Based on these findings, we selected Llama 3 for integration into the final chatbot.

\begin{table}[t]
\centering
\resizebox{\columnwidth}{!}{
  \begin{tabular}{l|c|c|c}
    \hline
    \textbf{Model} & \textbf{BLEU-1} & \textbf{BLEU-3} & \textbf{BLEURT} \\
    \hline
    \href{https://huggingface.co/openai-community/gpt2-medium}{\texttt{GPT-2 medium}} & 10.70 & 22.45 & -0.38 \\
    \href{https://huggingface.co/timinar/baby-llama-58m}{\texttt{BabyLlama}} & 10.30 & 22.27 & -0.42 \\
    \href{https://huggingface.co/google/flan-t5-base}{\texttt{FLAN-T5 base}} & \textbf{20.31} & 34.78 & -0.16 \\
    \href{https://huggingface.co/google/gemma-7b-it}{\texttt{Gemma 7B}} & 17.51 & 27.06 & -0.20 \\
    \href{https://huggingface.co/meta-llama/Meta-Llama-3-8B}{\texttt{Llama 3 8B}} & 19.16 & \textbf{36.48} & \textbf{-0.14} \\
    \href{https://huggingface.co/mistralai/Mistral-7B-Instruct-v0.2}{\texttt{Mistral 7B}} & 12.50 & 25.47 & -0.18 \\
    \hline
  \end{tabular}
}
\caption{Results from automatic evaluation of fine-tuned nutritional counselling models, using BLEU with 1 reference, BLEU with 3 references, and BLEURT. The best value for each metric is displayed in \textbf{bold}.}
\label{tab:experimentation-results}
\end{table}

We provide an example output from Llama 3, fine-tuned for nutritional counselling, in \Cref{tab:counselling-outputs}. Finally, we implement a conversation flow to guide users seeking help, as outlined in \Cref{fig:counselling-flow}.

\begin{table}[t]
\centering
\resizebox{\columnwidth}{!}{
    \begin{tabular}{p{0.5\textwidth}}
    \toprule
    \textbf{Struggle:} I eat more because of stress\\ \midrule
    \textbf{Reflection:} Are you saying that you eat more when you are feeling stressed?\\ \midrule
    \textbf{Comfort:} It's understandable to turn to food as a coping mechanism for stress.\\ \midrule
    \textbf{Reframing:} Something positive you could consider about this is that you are aware of the connection between stress and eating habits, and can take steps to address it.\\ \midrule
    \textbf{Suggestion:} Starting from tomorrow, you could try practicing mindfulness techniques such as deep breathing or meditation to help manage stress and reduce cravings for food.\\ \bottomrule
    \end{tabular}
}
\caption{Example outputs of each supportive text category from the fine-tuned Llama 3 model, in response to a user-given struggle.}
\label{tab:counselling-outputs}
\end{table}

\begin{table*}[ht]
\centering

\resizebox{\linewidth}{!}{\begin{tabular}{lccc|ccc|ccc|ccc|ccc|ccc}
\toprule
\textbf{Group} & \multicolumn{3}{c}{\textbf{kcal (\%)}} & \multicolumn{3}{c}{\textbf{Carbs (\%)}} & \multicolumn{3}{c}{\textbf{Protein (\%)}} & \multicolumn{3}{c}{\textbf{Fat (\%)}} & \multicolumn{3}{c}{\textbf{Sodium (\%)}} & \multicolumn{3}{c}{\textbf{Sugar (\%)}} \\
\midrule
               & $W1 (\downarrow)$ & $W7 (\downarrow)$ & $\Delta (\uparrow)$ & $W1 (\downarrow)$ & $W7 (\downarrow)$ & $\Delta (\uparrow)$ & $W1 (\downarrow)$ & $W7 (\downarrow)$ & $\Delta (\uparrow)$ &
                 $W1 (\downarrow)$ & $W7 (\downarrow)$ & $\Delta (\uparrow)$ & $W1 (\downarrow)$ & $W7 (\downarrow)$ & $\Delta (\uparrow)$ & $W1 (\downarrow)$ & $W7 (\downarrow)$ & $\Delta (\uparrow)$ \\
\midrule
\texttt{BASELINE}  & \textbf{21.96} & \textbf{21.32} &  \cellcolor{red!15}\textbf{0.64} & 34.95 & 32.50 &  2.45 & \textbf{35.41} & \textbf{30.87} & \cellcolor{green!15}\textbf{4.54} &
                 38.97 & 38.49 &  \cellcolor{green!15}\textbf{0.48} & \textbf{50.73} & 49.64 &  1.09 & 52.93 & 50.34 & \cellcolor{green!15}\textbf{2.58} \\
\texttt{REPHRASED} & 25.42 & 24.33 &  1.09 & \textbf{31.89} & 32.80 & \cellcolor{red!15}\textbf{-0.91} & 35.75 & 34.16 &  1.58 &
                 \textbf{35.67} & \textbf{37.11} & \cellcolor{red!15}\textbf{-1.44} & 51.96 & 61.07 & \cellcolor{red!15}\textbf{-9.11} & \textbf{49.03} & \textbf{49.34} & -0.31 \\
\texttt{FULL}      & 26.68 & 23.40 & \cellcolor{green!15}\textbf{3.28} & 37.56 & \textbf{31.22} & \cellcolor{green!15}\textbf{6.34} & 36.04 & 34.63 &  \cellcolor{red!15}\textbf{1.42} &
                 36.21 & 37.17 & -0.97 & 57.78 & 50.28 & \cellcolor{green!15}\textbf{7.49} & 49.95 & 53.11 & \cellcolor{red!15}\textbf{-3.16} \\
\bottomrule
\end{tabular}}
\caption{Group adherence to dietary goals. We report the absolute distance (\%) from calories and nutrient goals (on average per group). We compare differences ($\Delta$) in average between the first ($W1$) and last ($W7$) week of trial. {\colorbox{green!15}{\textbf{Best}}} and {\colorbox{red!15}{\textbf{worst}}} $\Delta$ highlighted.}
\label{tab:diet-improvement-table}
\vspace{1em}

\resizebox{\linewidth}{!}{\begin{tabular}{lcc|cc|cc|cc|cc|cc}
\toprule
\textbf{Group Comparison} & \multicolumn{2}{c|}{\textbf{kcal}} & \multicolumn{2}{c|}{\textbf{Carbs}} & \multicolumn{2}{c|}{\textbf{Fat}} & \multicolumn{2}{c|}{\textbf{Protein}} & \multicolumn{2}{c|}{\textbf{Sodium}} & \multicolumn{2}{c}{\textbf{Sugar}} \\ \midrule
                           & \textbf{Diff.} & \textbf{p-value} & \textbf{Diff.} & \textbf{p-value} & \textbf{Diff.} & \textbf{p-value} & \textbf{Diff.} & \textbf{p-value} & \textbf{Diff.} & \textbf{p-value} & \textbf{Diff.} & \textbf{p-value} \\
\midrule
\texttt{BASELINE} - \texttt{REPHRASED}      & -0.27 & 0.54 & 0.65  & 0.27 & -0.16 & 0.84 & -0.15 & 0.86 & 0.14  & 0.87 & 0.22  & 0.81 \\
\texttt{BASELINE} - \texttt{FULL}           & -0.38 & 0.38 & -0.69 & 0.23 & 0.17  & 0.84 & -0.31 & 0.71 & -0.88 & 0.29 & 0.02  & 0.98 \\
\texttt{REPHRASED} - \texttt{FULL}          & -0.11 & 0.79 & -1.34 & \textbf{0.02*} & 0.33  & 0.68 & -0.16 & 0.85 & -1.02 & 0.21 & -0.20 & 0.82 \\
\bottomrule
\end{tabular}}
\caption{Differences and p-values from the mixed-effects models comparing group pairs for energy and nutrients goals. We compare weekly changes per-metric for each group. Significant p-values are marked with an asterisk (*).}
\label{tab:diet-lmem}

\end{table*}

\section{Chatbot Trial}

We conducted a seven-week randomised controlled trial to test the chatbot's real-world impact from May to June 2024. Participants were recruited through social media, flyers, and direct outreach across multiple locations, resulting in a demographically diverse cohort. Exact demographics are outlined in \Cref{app:demographics}. Through an onboarding process, users were taught how to install the required apps, connect their accounts, access the chatbot, and use it. We successfully onboarded 87 participants, and 81 of them completed the full seven-week duration. Dropouts occurred across all study groups, primarily due to time constraints or lack of engagement. These early exits were balanced across conditions and did not compromise the integrity of the trial.

Participants were randomly assigned to one of three groups: \texttt{BASELINE} (n=26), with templated insights from the food diary only; \texttt{REPHRASED} (n=27), with LLM-rephrased responses; \texttt{FULL} (n=28), with LLM-rephrased messages and nutritional counselling. Throughout the trial, participants logged their meals daily via MyFitnessPal, engaged with the chatbot on Telegram five or more times per week, and completed a weekly emotional well-being questionnaire using the Positive and Negative Affect Schedule (PANAS) \citep{thompson2007development} (more details in \Cref{ssec:emotional-wellbeing}). Participation in the trial was incentivised through weekly online gift vouchers, with a doubled reward in the final week,
and adherence was monitored through MyFitnessPal logs and conversation history. While we encouraged regular participation, occasional lapses were tolerated as long as participants remained responsive and completed the weekly questionnaires.
Ethics and exact compensation details are provided in \Cref{app:ethics}.

During the trial, participant adherence was monitored using automated checks, including morning checks of the previous day's food log to identify incomplete or implausible entries, mid-afternoon reminders for missing entries, and evening nudges for inactive users (more details in \Cref{app:sanity-checks}). While nutritional counselling could be accessed anytime by the \texttt{FULL} group, the chatbot also actively offered it each Friday. At the end of each week, the chatbot provided participants with a link to the PANAS questionnaire to assess emotional well-being and released their weekly voucher upon completion. At the conclusion of the trial, offboarding had participants fill out a final feedback form tailored to their assigned study group. 

The trial faced several technical challenges, including a temporary disruption in the chatbot's access to MyFitnessPal data (beyond our control), a necessity to retrain the nutritional counselling model, and rare bugs that made the chatbot unusable for short periods of time (typically one hour or less, and fixed promptly). We discuss these in \Cref{app:tech-errors}.
 
\section{Results}

\subsection{Dietary outcome}

Participants in each group logged their daily dietary intake using MyFitnessPal, allowing us to evaluate adherence to the personal diet goals provided by the app. We focused on how the LLM-powered features in \texttt{REPHRASED} and \texttt{FULL} influenced intake behaviours compared to \texttt{BASELINE}. We first measure the absolute distance (\%) from user's intake goals (lower is better). Since different groups started at different distances, we consider the difference between the first and last weeks of the trial as a more objective measure. Using this approach, we avoid cases in which an improvement could be observed simply because the relevant group started, on average, closer to a specific goal.

\begin{table*}[t]
\centering
\begin{minipage}[t]{0.48\linewidth}
\centering
\small
\resizebox{\linewidth}{!}{\begin{tabular}{lrrr|rrr}
\toprule
\textbf{Group} & \multicolumn{3}{c}{\textbf{PA $(\uparrow)$}} & \multicolumn{3}{c}{\textbf{NA $(\downarrow)$}} \\
\midrule
               & $W1$ & $W7$ & $\Delta (\uparrow)$ & $W1$ & $W7$ & $\Delta (\downarrow)$\\
\midrule
\texttt{BASELINE}  &  \textbf{15.52} &  15.40 & \cellcolor{red!15}\textbf{-0.12} &   8.56 &  \textbf{8.44} & -0.12 \\
\texttt{REPHRASED} &  14.81 &  \textbf{16.77} &  \cellcolor{green!15}\textbf{1.96} &  10.69 &  9.77 & \cellcolor{green!15}\textbf{-0.92} \\
\texttt{FULL}      &  14.30 &  14.30 &  0.00 &   \textbf{8.26} &  8.78 &  \cellcolor{red!15}\textbf{0.52} \\
\bottomrule
\end{tabular}}
\caption{Average positive affect (PA) and negative affect (NA) scores in week 1 and week 7, and the delta between them. We compare difference ($\Delta$) in average between the last ($W7$) and first ($W1$) week of trial. {\colorbox{green!15}{\textbf{Best}}} and {\colorbox{red!15}{\textbf{worst}}} $\Delta$ highlighted.}
\label{tab:panas-deltas}
\end{minipage}
\hspace{0.4cm} 
\begin{minipage}[t]{0.48\linewidth}
\centering
\small
\resizebox{\linewidth}{!}{\begin{tabular}{lcc|cc}
\toprule
\textbf{Group} & \multicolumn{2}{c|}{\textbf{PA}} & \multicolumn{2}{c}{\textbf{NA}} \\ \midrule
                           & \textbf{Diff.} & \textbf{p-value} & \textbf{Diff.} & \textbf{p-value} \\
\midrule
\texttt{BASELINE} - \texttt{REPHRASED}      & 0.14 & 0.34 & -0.13 & 0.47 \\
\texttt{BASELINE} - \texttt{FULL}           & -0.05 & 0.73 & 0.18  & 0.31 \\
\texttt{REPHRASED} - \texttt{FULL}          & -0.19 & 0.20 & 0.31  & 0.08 \\
\bottomrule
\end{tabular}}
\caption{Differences and p-values from the mixed-effects models for the PANAS scores.  We compare weekly changes per-score for each group. Significant p-values are marked with an asterisk (*).}
\label{tab:panas-lmem}

\end{minipage}%
\end{table*}

An initial look at results (\Cref{tab:diet-improvement-table}) looks promising: for the three metrics where \texttt{FULL} shows the greatest improvement (kcal, carbs and sodium), the values proved 2x-7x more than that of \texttt{BASELINE}. \texttt{REPHRASED}, on the other hand, never showed a greater improvement than the other groups and actually showed the worst result out of the three metrics. 

However, the improvement values fall within a very small range: we see cases where the ``best'' group shows less than a 1\% improvement, and even the greatest values do not go past $\sim$7.5\%. Therefore, we check for any significance in these results through a linear mixed-effects model (\Cref{tab:diet-lmem}). Here, the narrative changes: we find no significance except for a group-by-time interaction for carbohydrate adherence in \texttt{FULL} compared to \texttt{REPHRASED}, hinting at an improved alignment with carbohydrate targets over time. Considering the lack of significance for any other measure, we deem this to be insufficient evidence of the benefits provided by LLM-based features. We report further insights and visualisation on dietary outcomes in \Cref{app:diet-by-goal}.

\subsection{Emotional Well-being}
\label{ssec:emotional-wellbeing}

The PANAS questionnaire is a validated self-assessment tool to measure emotional affect. Participants are asked to rate specific emotions on a scale of one to five, based on the extent they felt them in the past week. From this, PANAS returns two independent scores: positive affect (PA, higher is better) and negative affect (NA, lower is better). To analyse the effect of LLM-based features on emotional state, we monitor the weekly change in both PA and NA. Ideally, we would expect a noticeable improvement in \texttt{FULL}, since this group had access to the nutritional counselling feature, allowing them to receive tailored empathetic support. 

The overall results (\Cref{tab:panas-deltas}) do not show particularly evident trends. \texttt{FULL} had no change in PA, and a negligible worsening in NA. In contrast, \texttt{REPHRASED} displayed the largest PA and NA improvements. \texttt{BASELINE} exhibited opposite trends for the two measures, with a slight decline in PA but a similarly small improvement in NA.

Again, the observed changes are relatively small. The fact that the participants from \texttt{REPHRASED} demonstrated the greatest improvements to emotional wellbeing out of all groups contradicts our hypotheses. This pattern would suggest that rephrasing alone noticeably boosts emotional wellbeing, while nutritional counselling has the opposite effect. As before, we run a linear mixed-effects model \Cref{tab:panas-lmem} to inspect whether any of these changes are statistically significant. The model did not identify any significant effects, including for \texttt{REPHRASED}. We further analyse the individual emotions targeted by PANAS by breaking down the scores (more information in \Cref{app:panas}). However, no emerging trend or significance was found.

\begin{table*}[ht]
\centering
\resizebox{0.75\linewidth}{!}{\begin{tabular}{lrrr|rrr|rrr}
\toprule
\textbf{Group} & \multicolumn{3}{c}{\textbf{Interactions}} & \multicolumn{3}{c}{\textbf{Conversations}} & \multicolumn{3}{c}{\textbf{Days}} \\
\midrule
& $W1 (\uparrow)$ & $W7 (\uparrow)$ & $\Delta (\uparrow)$ & $W1 (\uparrow)$ & $W7 (\uparrow)$ & $\Delta (\uparrow)$ & $W1 (\uparrow)$ & $W7 (\uparrow)$ & $\Delta (\uparrow)$\\
\midrule
\texttt{BASELINE}  &  23.15 &   9.77 & \cellcolor{green!15}\textbf{-13.38} &  \textbf{8.08} &  5.19 & \cellcolor{red!15}\textbf{-2.88} &  \textbf{5.31} &  4.38 & \cellcolor{red!15}\textbf{-0.92} \\
\texttt{REPHRASED} &  24.30 &  10.22 & -14.07 &  7.74 &  5.63 & -2.11 &  5.30 &  4.74 & -0.56 \\
\texttt{FULL}      &  \textbf{27.75} &  \textbf{13.39} & \cellcolor{red!15}\textbf{-14.36} &  7.54 &  \textbf{6.25} & \cellcolor{green!15}\textbf{-1.29} &  4.93 &  \textbf{5.14} &  \cellcolor{green!15}\textbf{0.21} \\
\bottomrule
\end{tabular}}
\caption{The count of interactions, conversations (interactions with less than 5 minutes in between), and interaction days across each group. We compare difference ($\Delta$) in average between the first ($W1$) and last ($W7$) week of trial. {\colorbox{green!15}{\textbf{Best}}} and {\colorbox{red!15}{\textbf{worst}}} $\Delta$ highlighted.}
\label{tab:engagement-deltas}
\end{table*}

\begin{table*}[ht]
\centering
\small
\begin{tabular}{lcc|cc|cc}
\toprule
\textbf{Group Comparison} & \multicolumn{2}{c|}{\textbf{Interactions}} & \multicolumn{2}{c|}{\textbf{Conversations}} & \multicolumn{2}{c}{\textbf{Days}} \\
                           & \textbf{Diff.} & \textbf{p-value} & \textbf{Diff.} & \textbf{p-value} & \textbf{Diff.} & \textbf{p-value} \\
\midrule
\texttt{BASELINE - REPHRASED}      & -0.02 & 0.96 & 0.08  & 0.53 & 0.07  & 0.26 \\
\texttt{BASELINE - FULL}           & -0.22 & 0.67 & 0.21  & 0.11 & 0.14  & \textbf{0.02*} \\
\texttt{REPHRASED - FULL}          & -0.20 & 0.69 & 0.13  & 0.34 & 0.07  & 0.25 \\
\bottomrule
\end{tabular}
\caption{Differences in engagement metrics (by count) and p-values from the mixed-effects models. Significant p-values are marked with an asterisk (*).}
\label{tab:engagement-lmem}
\end{table*}

\subsection{User Engagement}

To investigate changes in engagement, we calculated the number of interactions (individual messages from the user), conversations (a sequence of interactions with responses within five minutes of each other), and days of chatbot use over the seven-week intervention. We report more detailed engagement metrics in~\Cref{app:engagement}. The utility of LLM-based functions here cause a noticeable increase in the above metrics, indicating that the \texttt{FULL} and \texttt{REPHRASED} groups spent more time interacting with the chatbot.

Our results (\Cref{tab:engagement-deltas}) show a consistent decline in all engagement metrics over time, regardless of the group. This indicates a natural decrease in user engagement over the weeks. \texttt{FULL} consistently showed higher interaction levels, likely driven by the additional nutritional counselling functionality and the weekly direct prompt encouraging feature use. In contrast, \texttt{REPHRASED} did not exhibit notable differences in engagement compared to \texttt{BASELINE}. Using a linear mixed-effects model, we find only one significant difference: participants in \texttt{FULL} spent significantly more days interacting with the chatbot than those in \texttt{BASELINE}. This suggests that people in \texttt{FULL} spent significantly more days interacting with the chatbot. Given the additional promotion of the ``advice'' feature in \texttt{FULL}, the lack of significant differences across other metrics, and the overall downward trend, we do not consider this sufficient evidence to support consistent benefits of LLM-based features.

\subsection{User Feedback}

User feedback collected at the end of the trial provided insight into the perceived strengths and weaknesses of the chatbot. Over 39\% of participants from all groups judged the visualisations accompanying the advanced insights as helpful for understanding their nutritional data. Users also appreciated the food diary reminders, time-period comparisons, nutritional breakdowns and the chatbot's conversational tone and ease of use.

However, many participants (50\% of \texttt{FULL}, 59\% of \texttt{REPHRASED}, and 32\% of \texttt{BASELINE}) reported difficulties with the chatbot's NLU module, specifically when typos were present or phrasing deviated from expected patterns. Some also struggled with fully using the chatbot despite the manual provided. An additional concern was the chatbot's limited functionality when compared to tools like ChatGPT. This reflects users' increasing expectations shaped by open-ended LLMs, although in healthcare contexts, stricter safety and reliability constraints apply. Although hallucinations were rarely reported, several users were frustrated by vague insights or inaccuracies stemming from the food diary data, and requested more tailored and concrete suggestions. In particular, users wanted meal recommendations, which we had to exclude because of safety compliance.

Regarding nutritional counselling, some participants appreciated the supportive tone, while others criticised the advice as overly generic, unhelpful, or even counterproductive. Some users felt that the attempt to offer comfort detracted from the clarity and speed of getting useful responses. One participant specifically criticised the open-ended nature of the advice, noting that statements like, ``You could consider that snacking can be a way to fuel your body and give it the energy it needs,'' risk inadvertently endorsing unhealthy habits. These issues align with the results reported by the dataset authors~\citep{balloccu-etal-2024-ask}.

\section{Conclusion}

As of today, there is an increasing trust in applying LLMs to healthcare and its related sub-domains, including nutrition. However, these models are typically evaluated intrinsically, on unrealistic or simulated tasks, and mostly relying on metrics. In this work, we present the first objective evaluation of the real-world impact of LLMs in nutrition. We integrated them into a chatbot for diet coaching and nutritional counselling, and ran a seven-week RCT on a population of 81 participants. 

Our results quickly point out the limits of intrinsic LLM evaluation. We found no consistent improvement across our metrics. Although some statistically significant improvements emerged, these appear spurious in the larger context of our trial. Based on our results, these models are unable to effectively promote dietary adherence, reduce the emotional load of dieting, provide empathetic help through counselling, or simply boost user engagement.

We conclude that, at the current time, there is no evident benefit in applying LLMs to nutrition in the setup we investigated. While our findings are specific to this domain and trial configuration, they serve as a step towards real-world evaluation in healthcare, highlighting how LLMs might (or might not) affect user outcomes in chatbot-driven nutritional counselling. Our overall research underscores the importance of critical evaluation of LLMs in health-focused applications. Although these models are often heralded for their potential to deliver dynamic and personalised interactions, our findings caution against adoption without rigorous real-world validation. We hope our results will shed light on the need for real-world assessments beyond benchmarks.

\section{Limitations} \label{sec:limitations}

We faced several limitations in this work that highlight areas for improvement in the development and deployment of the diet-coaching chatbot. One significant issue was the natural language understanding component of the chatbot. Despite training Rasa's pipeline on a sufficient number of in-domain examples, the chatbot struggled with the varied user inputs, particularly in processing different date formats, which the chatbot relied on an entity parser to extract. Users often requested insights in diverse ways and with various date notations, leading to occasional misunderstandings or failures to deliver the requested insights. This challenge impacted the chatbot's interactions, and the overall user experience.

Furthermore, the trial's reliance on a task-specific chatbot exposes its limited adaptability when compared to open-domain models that the general public is usually exposed to, such as ChatGPT. Unfortunately, in sensitive domains, task-specific design is mandatory, as it allows more focused and controlled interventions. An RCT with an open-ended chatbot, able to answer any query from the user would have been too dangerous to be allowed. Because of this, we had to restrict our chatbot’s ability to handle diverse conversational contexts or unexpected queries, which are strengths of modern open-domain models. This highlights a trade-off between specialisation and adaptability.

Another limitation stemmed from the dataset used to train the nutritional counselling feature. \texttt{HAI-Coaching}, while expert-annotated as safe, also contains overly generic advice. Consequently, the chatbot's recommendations often lacked the depth and personalisation necessary for more impactful dietary advice. This issue was further aggravated by the lack of integration between user-reported struggles and their submitted food diaries. Although the food diary data were collected and processed separately for analysis of the trial outcomes, the chatbot did not use them to tailor its advice. This omission, which several trial participants pointed out, could further enhance the nutritional counselling feature.

Additionally, our implementation relied on relatively smaller models rather than larger, more powerful models. This choice was primarily driven by hardware limitations and the practical necessity for fast inference times; the chatbot needed to handle concurrent interactions with all trial participants and provide responses within a few seconds to maintain conversational flow. It remains possible that larger models, with greater capacity and more nuanced language understanding, might have delivered improved counselling quality or engagement.

Finally, as this is the first RCT evaluating LLM-based nutritional counselling, there is limited prior evidence on the timescale over which such interventions might influence dietary outcomes. Unlike previous trials using template-based messages, there are no directly comparable studies to guide expected effect onset. Consequently, the seven-week duration of our trial was chosen pragmatically, constrained by available funding and resources rather than established guidance. Future work could investigate optimal intervention duration, potentially informed by emerging evidence or longitudinal studies, to better contextualise the timing and magnitude of effects.

\section{Acknowledgements}

This work has been funded by the EC in the H2020 Marie Skłodowska-Curie Phil\-Humans project (contract no.\ 812882) and the European Research Council (Grant agreement No.\ 101039303 NG-NLG). It is further supported by the DYNAMIC center, funded by the LOEWE program of the Hessian Ministry of Science and Arts (grant number: LOEWE1/16/519/03/09.001(0009)/98).

\bibliography{custom}

\appendix
\renewcommand\thefigure{\thesection.\arabic{figure}}
\renewcommand\thetable{\thesection.\arabic{table}}

\newpage
~\\
\newpage

\section{Trial Checks}
\label{app:sanity-checks}
Throughout the duration of the trial, the chatbot performed several daily checks on participants to ensure and encourage task adherence:
\begin{itemize}
    \item At 10am, the chatbot checked for any \textbf{abnormalities} in the previous day's food logs, considering a combination of objective measures and heuristic decisions. The abnormalities found could be to do with mistakes in logging, or diary incompleteness. If it noticed an abnormality, the chatbot sent a message asking the user to double check their food diary on MyFitnessPal. These checks investigated whether the user's food diary from the previous day:
    \begin{itemize}
        \item was empty;
        \item consisted of less than half their calorie goal;
        \item consisted of more than twice their calorie goal;
        \item had less than four food items recorded;
        \item contained a particular food item with a recorded amount of more than one kilogram,  two litres, or six cups.
    \end{itemize}
    \item At 4pm, the chatbot sent a reminder to any user with an \textbf{empty diary} that day, i.e. anyone who had yet to log a food item.
    \item At 6pm, if the user had \textbf{not interacted} with the chatbot in the last \textbf{36 hours}, the chatbot encouraged them to initiate a conversation.
\end{itemize}

\section{Technical Details}
\label{app:tech-details}
\setcounter{figure}{0}
\setcounter{table}{0}
For fine-tuning the nutritional counselling models, we had access to NVIDIA A40 GPUs with 48GB of VRAM. We applied 4-bit quantisation to optimise memory usage and enable efficient training of the larger of the chosen models (Gemma 7B, Llama 3 8B, and Mistral 7B). The prompts are shown in \Cref{tab:finetuning-prompts} and the training configurations are outlined in \Cref{tab:train-parameters}. We monitored validation loss during training and selected the best-performing checkpoints. Batch sizes and training durations were adjusted to match the capabilities of the available GPUs.

\begin{table*}[!htb]
    \centering
    \begin{tabular}{p{0.15\linewidth} | p{0.72\linewidth}}\hline
        \textbf{Text} & \textbf{Prompts} \\ \hline
        \textbf{REFL}&
        You are an expert dietitian. Below is a struggle your client is experiencing. Summarize what the problem is about or infer what they mean. Do not assume their feelings.
        \newline\newline
        \#\#\# Struggle:\newline
        \texttt{[STRUGGLE]}
        \newline\newline
        \#\#\# Reflection: \vspace{1mm}\\
        \hline 
        \textbf{COMF}&
        You are an expert dietitian. Below is a struggle your client is experiencing. Tell them that the situation is not unrecoverable, normalize the situation or make them feel understood. Do not normalize dangerous behaviours in a way that explicitly encourages your client to commit them.
        \newline\newline
        \#\#\# Struggle:\newline
        \texttt{[STRUGGLE]}
        \newline\newline
        \#\#\# Comfort: \vspace{1mm}\\
        \hline 
        \textbf{REFR}&
        You are an expert dietitian. Below is a struggle your client is experiencing. Show a benefit to the struggle that they did not consider or find something about the struggle to be grateful for.
        \newline\newline
        \#\#\# Struggle:\newline
        \texttt{[STRUGGLE]}
        \newline\newline
        \#\#\# Reframing: \vspace{1mm}\\
        \hline 
        \textbf{SUGG}&
        You are an expert dietitian. Below is a struggle your client is experiencing. Tell the person how to change their habit to improve or suggest an alternative helpful activity.
        \newline\newline
        \#\#\# Struggle:\newline
        \texttt{[STRUGGLE]}
        \newline\newline
        \#\#\# Suggestion: \vspace{1mm}\\
        \hline
    \end{tabular}
    \caption{The instruction prompts used to fine-tune the instruction-tuned models.}
    \label{tab:finetuning-prompts}
\end{table*}

Towards LLM-based rephrasing, we selected the models according to our technical constraints, as with the nutritional counselling models, but with extra consideration of the inference speeds. This was initially a major limitation despite a simpler prompting approach on the machine deploying the chatbot. Running full models on the available Quadro RTX 5000 GPU resulted in average decoding times of over 30 seconds per message—too slow for a real-time chatbot. We explored several optimisation strategies, including 4-bit quantisation, KV-caching, \verb|torch.compile|, flash attention. While most techniques yielded minimal improvements, combining the 4-bit quantised Llama 3 with the Ollama API drastically reduced decoding time to around 1.1 seconds per message, making real-time deployment viable. Results from an initial prompt (\Cref{fig:rephrasing-prompt-init}) on 19 example messages with varying intents and complexity are outlined in \Cref{tab:decoding-time}. Among the three models, Mistral was quickly ruled out due to inconsistent outputs and occasional language mixing. Gemma performed better, but tended to omit key information and exhibited slower inference times. Llama 3 8B emerged as the most consistent, producing diverse and accurate rephrasings, and was selected as the production model.

\begin{table}[ht]
    \centering
    \resizebox{\columnwidth}{!}{
    \begin{tabular}{l|cccc}\hline
        \textbf{Model} & \textbf{Batch} & \textbf{Epochs} & \textbf{LR} & \textbf{Optimiser} \\ \hline
        \href{https://huggingface.co/openai-community/gpt2}{\texttt{GPT-2 small}} & 8 & 10 & 5e-5 & \texttt{AdamW} \\
        \href{https://huggingface.co/openai-community/gpt2-medium}{\texttt{GPT-2 medium}} & 4 & 10 & 5e-5 & \texttt{AdamW} \\
        \href{https://huggingface.co/timinar/baby-llama-58m}{\texttt{BabyLlama}} & 8 & 20 & 5e-5 & \texttt{AdamW} \\
        \href{https://huggingface.co/google/flan-t5-base}{\texttt{FLAN-T5 base}} & 8 & 30 & 1e-4 & \texttt{AdamW} \\
        \href{https://huggingface.co/google/gemma-7b-it}{\texttt{Gemma 2B}} & 2 & 3 & 2e-4 & \texttt{paged\_adamw\_8bit} \\
        \href{https://huggingface.co/google/gemma-7b-it}{\texttt{Gemma 7B}} & 2 & 3 & 2e-4 & \texttt{paged\_adamw\_8bit} \\
        \href{https://huggingface.co/mistralai/Mistral-7B-Instruct-v0.2}{\texttt{Mistral 7B}} & 2 & 3 & 2e-4 & \texttt{paged\_adamw\_8bit} \\
        \href{https://huggingface.co/meta-llama/Meta-Llama-3-8B}{\texttt{Llama 3 8B}} & 2 & 3 & 2e-4 & \texttt{paged\_adamw\_8bit} \\ \hline
    \end{tabular}}
    \caption{Parameters for the fine-tuning of the nutritional counselling models. LR = learning rate.}
    \label{tab:train-parameters}
\end{table}

\begin{table}[ht]
    \centering
    \begin{tabular}{l|c|c|c}
            \hline
            \textbf{Technique} & \textbf{Llama 3} & \textbf{Gemma} & \textbf{Mistral}\\
            \hline
            Full model  & 29  & 48  & 31\\
            4-bit model & 6.3 & 7.6 & 11\\
            Unsloth     & 34  & --- & --- \\
            Ollama      & 1.1 & 1.1 & ---\\
            \hline
    \end{tabular}
    \caption{Average decoding time in seconds per example for each of the 19 example messages rephrased by the models using various methods.}
    \label{tab:decoding-time}
\end{table}

\section{Rephrasing Evaluation}
\label{app:rephrasing-eval}
\setcounter{figure}{0}
\setcounter{table}{0}

\begin{table}[ht]
    \centering
    \begin{tabular}{c|c|c}
        \hline
        \textbf{Text pair} & \textbf{Preferred} & \textbf{More natural} \\
        \hline
        1 & 70\% & 85\% \\
        2 & 75\% & 85\% \\
        3 & 65\% & 65\% \\
        4 & 55\% & 65\% \\
        5 & 60\% & 55\% \\
        6 & 75\% & 90\% \\
        7 & 55\% & 60\% \\
        8 & 65\% & 75\% \\
        9 & 60\% & 75\% \\
        10 & 85\% & 80\% \\
        11 & 55\% & 65\% \\
        \hline
    \end{tabular}
    \caption{Results of the human evaluation of the rephrased responses in comparison to the templated responses in terms of proportion of participants in favour of the rephrased response.}
    \label{tab:prolific-preference-results}
\end{table}

\begin{table*}[ht]
    \centering
    \begin{tabular}{c|c|p{0.6\linewidth}}
        \hline
        {} & \textbf{Texts} & \textbf{Explanation} \\
        \hline
        1 & 1 & The “on average” section had the numbers reversed. \\
        2 & 1 & One has a 40\% deficit and one has a 60\% one \\
        {} & 9 & One asks to be let know what changed one does not \\
        3 & 4 & [The templated response] states 'to see how different foods contribute to your intake' but [the rephrased response] states 'foods that contributed to your daily intake' \\
        {} & 6 & [The templated response] is referring to safety but [the rephrased response] is referring to the accuracy of the results \\
        4 & 9 & ----- \\
        5 & 1 & They differed in the average protein that the user consumed, [the rephrased response] claimed they only hit 40\% of their protein goal while [the templated response] claimed that they hit 60\%. \\
        6 & 1 & Numbers juxtapositioned \\
        \hline
    \end{tabular}
    \caption{Results of the human evaluation of the rephrased responses in comparison to the templated responses in terms of the reporting of differences in meaning.}
    \label{tab:prolific-meaning-results}
\end{table*}

In order to verify the suitability of the rephrasing model, we ran an intrinsic evaluation with human crowd workers to compare the templated responses against the rephrased ones. In this experiment, our objective was to evaluate the preference of chatbot users towards the original templated responses of the baseline chatbot or the responses that have been rephrased by the prompted model. Furthermore, we aimed to determine which response sounded more natural to users and whether users could distinguish any difference in meaning between the two texts. Only a few semantic mismatches were reported, typically due to numerical misinterpretation in isolated cases.

The evaluation took the form of an annotation task that presented a random sample of 12 pairs of templated responses and rephrased outputs, with accompanying conversational context, covering a diverse range of user queries and scenarios. These text pairs are included in our repository. The participants were shown these text pairs as a sequence of twelve questions, including an attention question designed to make sure that they were completing the task according to the instructions. To each presented text pair, they had to answer three sub-questions:
\begin{enumerate}
    \item Which response do you prefer?
    \item Which response sounds more natural?
    \item Do both responses have the same meaning? \textit{If the participant responded with "No", they could optionally provide a reason.}
\end{enumerate}

Participants of this annotation task were sourced on Prolific (\href{https://www.prolific.com}{https://www.prolific.com}), under conditions that they were primarily an English speaker. In total, 23 crowd workers on the platform completed the task, of which 20 passed the attention question to be considered in our analysis.

The results of the task are displayed in \Cref{tab:prolific-preference-results,tab:prolific-meaning-results}. Overall, the findings suggest that while participants generally preferred and found the rephrased responses more natural compared to the templated responses, the preference was not overwhelmingly strong, with some text pairs showing a narrower margin of preference. While the templated outputs featured more structured formatting through the use of newlines and bolding, the rephrased outputs leaned toward a more conversational style, often incorporating emojis as instructed. This difference in presentation may have influenced participant preferences and contributed to the higher perceived naturalness of the rephrased responses.

\section{Ethics Details}
\label{app:ethics}
\setcounter{figure}{0}
\setcounter{table}{0}

Ethical approval was obtained from the University of Aberdeen as well as ethical advisors at Charles University, and informed consent was obtained twice from all trial participants during onboarding: once during registration and again at onboarding, ensuring participants fully understood the study tasks (after group assignment) and data usage. Participants had the right to withdraw at any point before data analysis began, and any data from those who withdrew early was excluded from analysis. As compensation, participants were given online gift vouchers (Alza or Amazon) worth €8, AU\$13, or 200CZK for each completed week of the first six weeks of the experiment, and €16, AU\$26, or 400CZK for the seventh (and final) week of the experiment, if completed.

\section{Technical Challenges During the Trial}
\label{app:tech-errors}
\setcounter{figure}{0}
\setcounter{table}{0}

During the initial setup of the model for the user trial, the model was trained on the first safe supportive text, prioritising the candidates provided by experts. However, part-way through the user trial, we identified an oversight: we had inadvertently ignored many other safe candidates in training that make up about seven times the number of training examples in the initial fine-tuning. That is, in the original model, a struggle was trained with a single corresponding text category, but, in fact, there are up to ten generated candidates and up to three expert-provided texts that can be used for training. To address this, the model was immediately retrained on the full set of safe candidates and swapped out with the original model serving the chatbot in the user trial in time for the fifth week since the beginning of the trial.

This approach provided the opportunity to compare the performance between the original and updated models. At the conclusion of the trial, we asked users if they observed any difference in the performance of the nutritional counselling model since the trial start. About 28\% of participants (n=8) agreed that they noticed an improvement, while 48\% (n=13) provided a neutral response, indicating no significant change in their experience. The remaining 24\% (n=7) disagreed. These results imply that the additional training data may not have led to substantial improvements in model performance. However, it is important to consider that the declining use of the ``advice'' feature over time, like the general decrease in chatbot interactions, may have influenced these perceptions.

The trial also faced several other technical challenges. The most significant was a temporary disruption in access to MyFitnessPal data between Weeks 4 and 5 (June 18–21), which prevented the chatbot from delivering personalised dietary insights. During this time, the nutritional counselling feature remained active. Users were informed of the issue and encouraged to continue logging meals independently. Full functionality was restored by June 21, and daily interaction requirements were relaxed to accommodate the interruption. Other disruptions that occurred were infrequent and resolved promptly.

\section{Trial Demographics}
\label{app:demographics}
\setcounter{figure}{0}
\setcounter{table}{0}

\begin{figure*}[ht]
    \includegraphics[width=\linewidth]{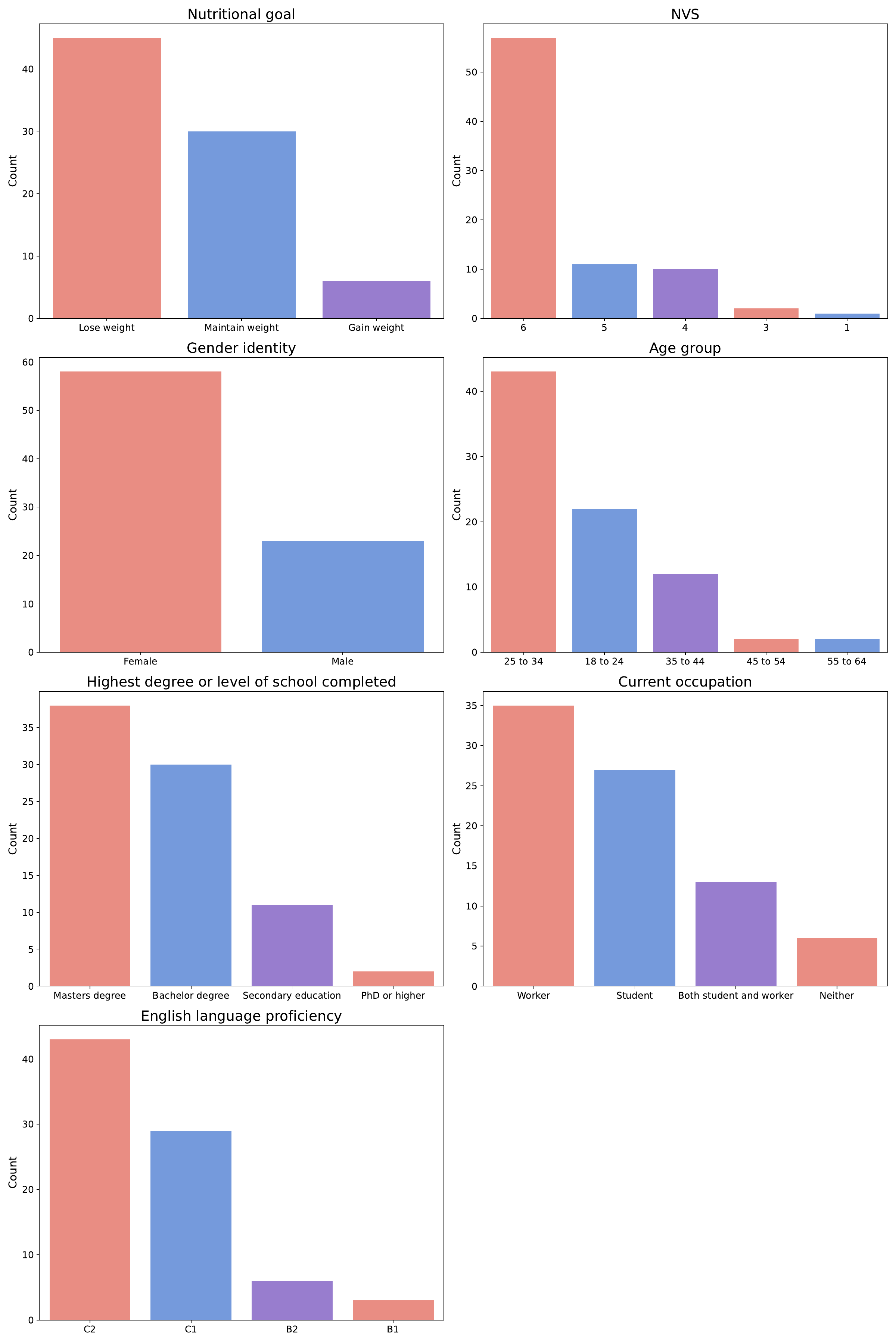}
    \caption{Graphics showing collected demographic data of participants (part 1).}
    \label{fig:demographics1}
\end{figure*}

\begin{figure*}[ht]
    \centering
    \includegraphics[width=\linewidth]{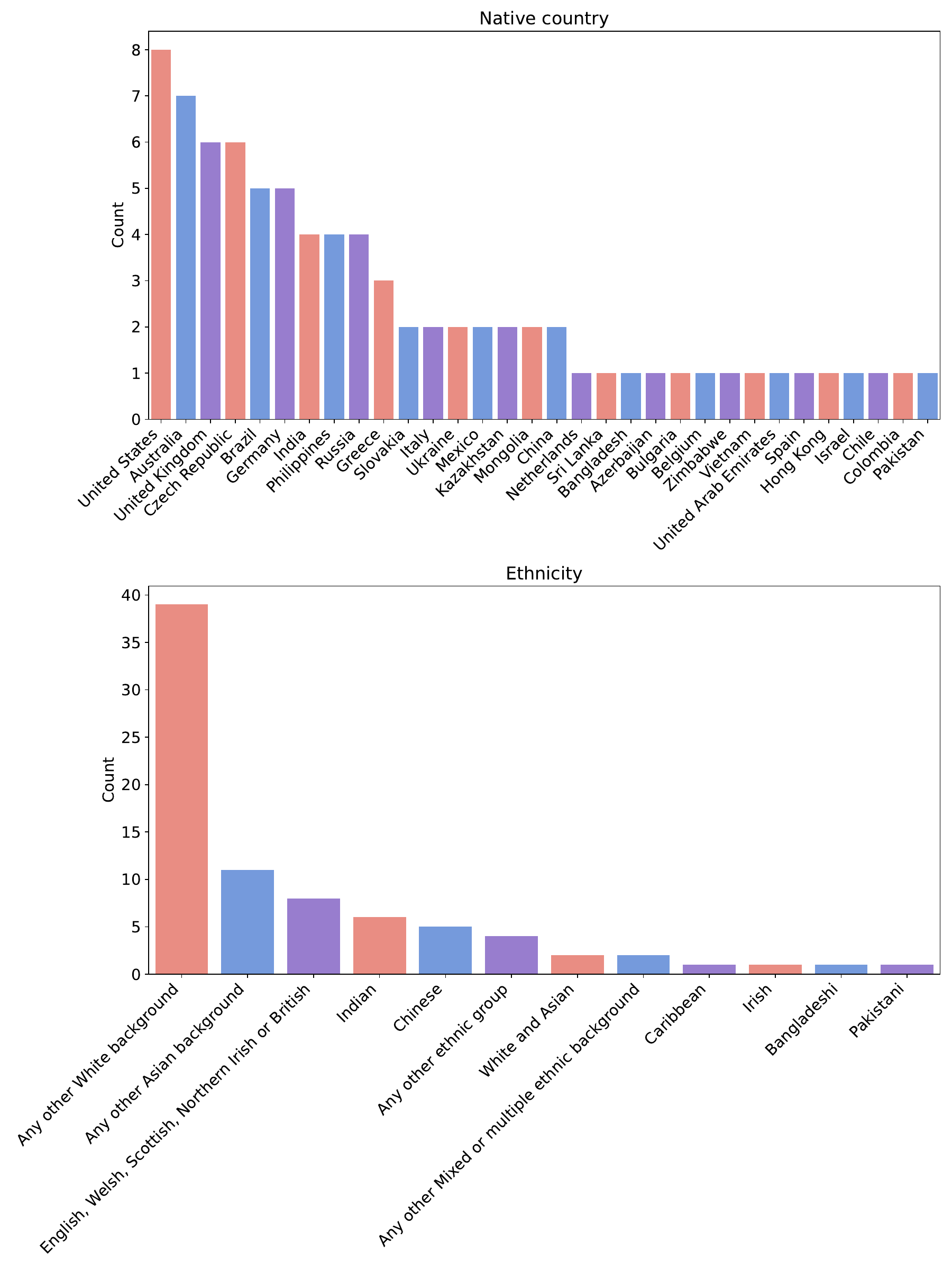}
    \caption{Graphics showing collected demographic data of participants (part 2).}
    \label{fig:demographics2}
\end{figure*}

Demographic data collected at registration included age, gender identity, educational background, occupation, ethnicity, native country, English proficiency, and nutritional literacy captured through Pfizer's NVS questionnaire \citep{Weiss2005}. The collected statistics are illustrated across \Cref{fig:demographics1,fig:demographics2}.

 The final study group was predominantly female, well-educated, ethnically diverse, and with adequate English skills and nutritional literacy for engaging with the chatbot and interpreting dietary insights.

\section{Trial Dietary Results by Weight Goal}
\label{app:diet-by-goal}
\setcounter{figure}{0}
\setcounter{table}{0}

\begin{figure*}[ht]
    \centering
    \includegraphics[width=1\linewidth]{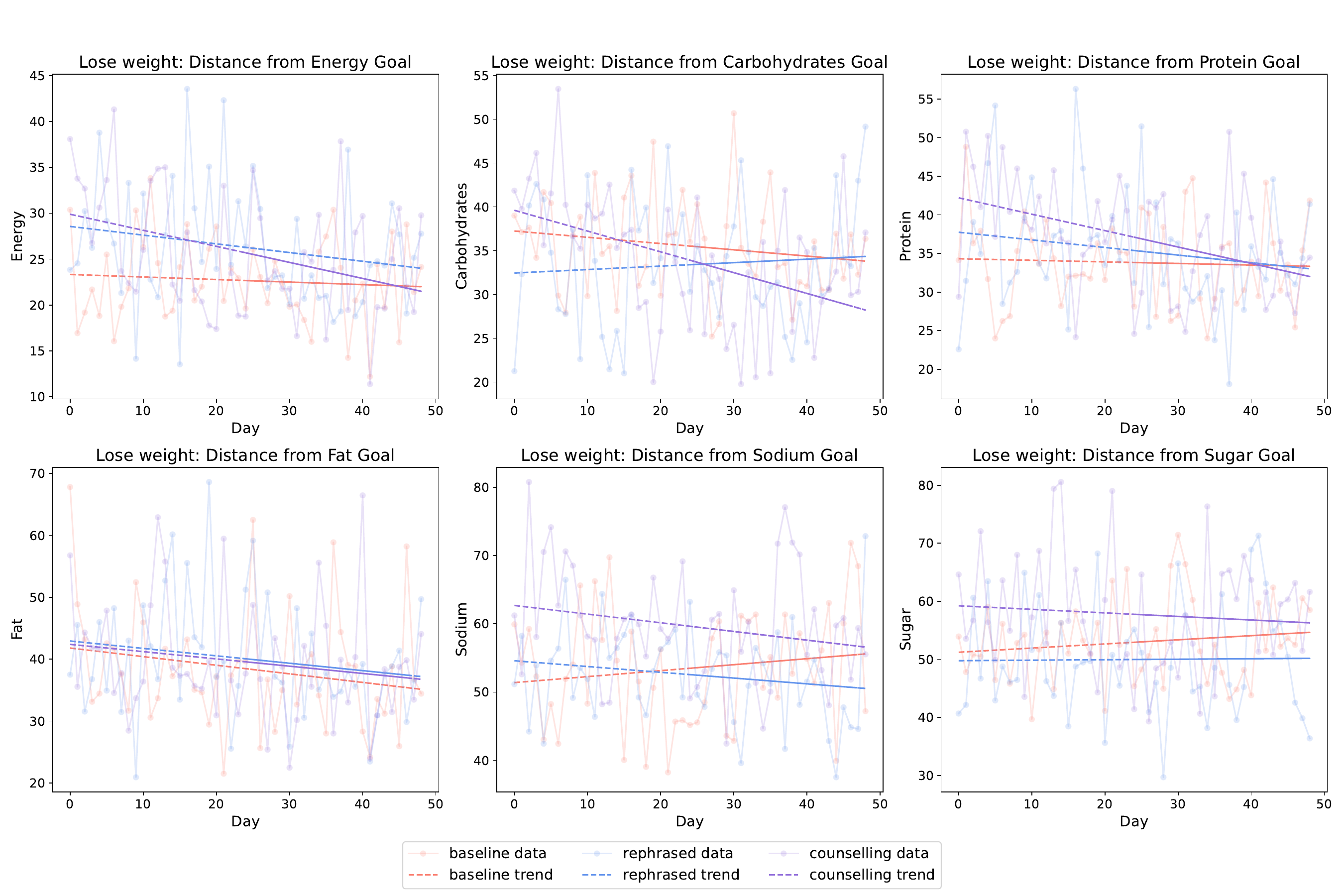}
    \caption{Absolute percentage distance from MyFitnessPal intake goals for participants aiming to lose weight. LW}
    \label{fig:mfp-lose-distance}
\end{figure*}

\begin{figure*}[ht]
    \includegraphics[width=1\linewidth]{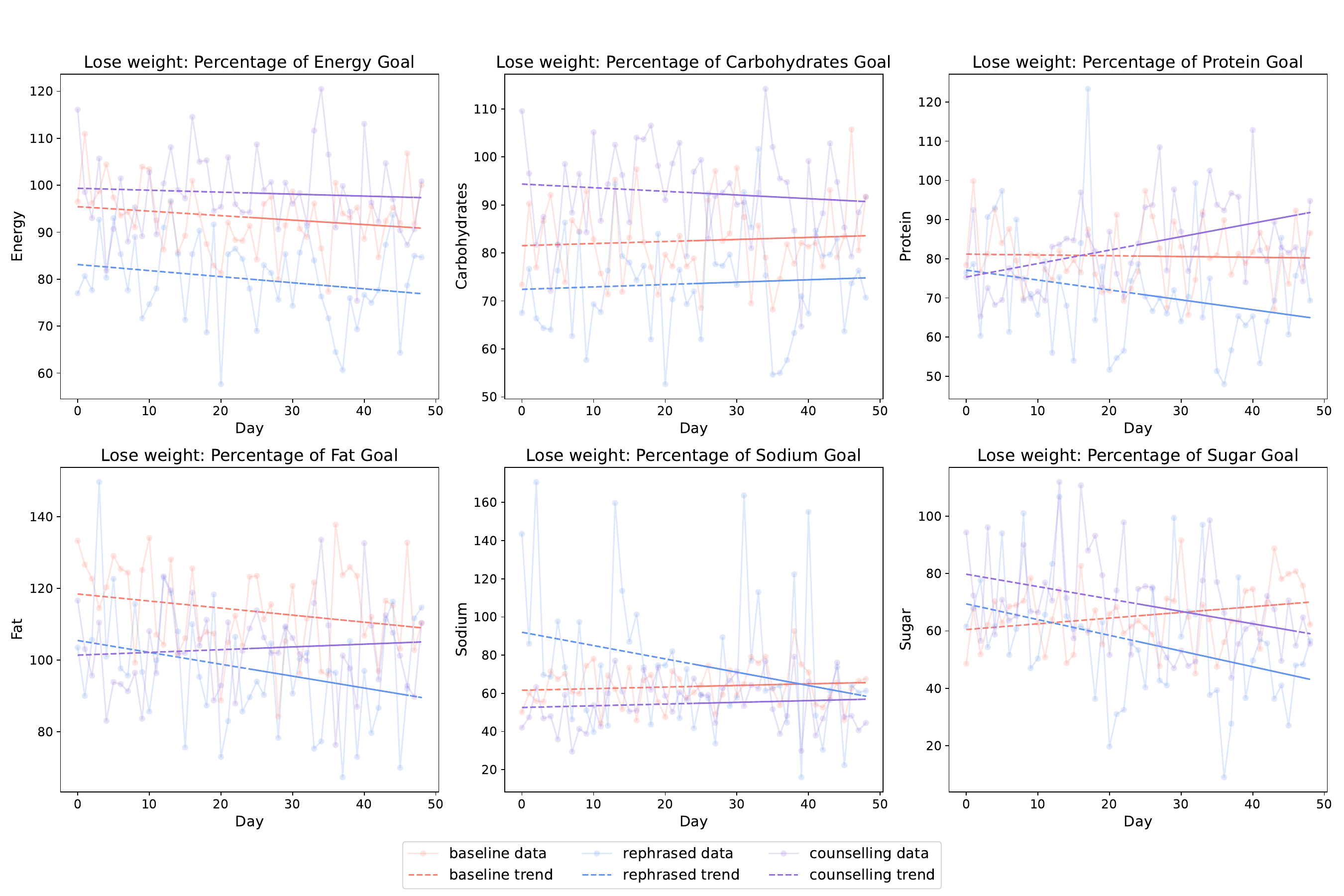}
    \caption{Goal percentage of MyFitnessPal intake goals for participants aiming to lose weight.}
    \label{fig:mfp-lose-percentage}
\end{figure*}

\begin{figure*}[ht]
    \centering
    \includegraphics[width=1\linewidth]{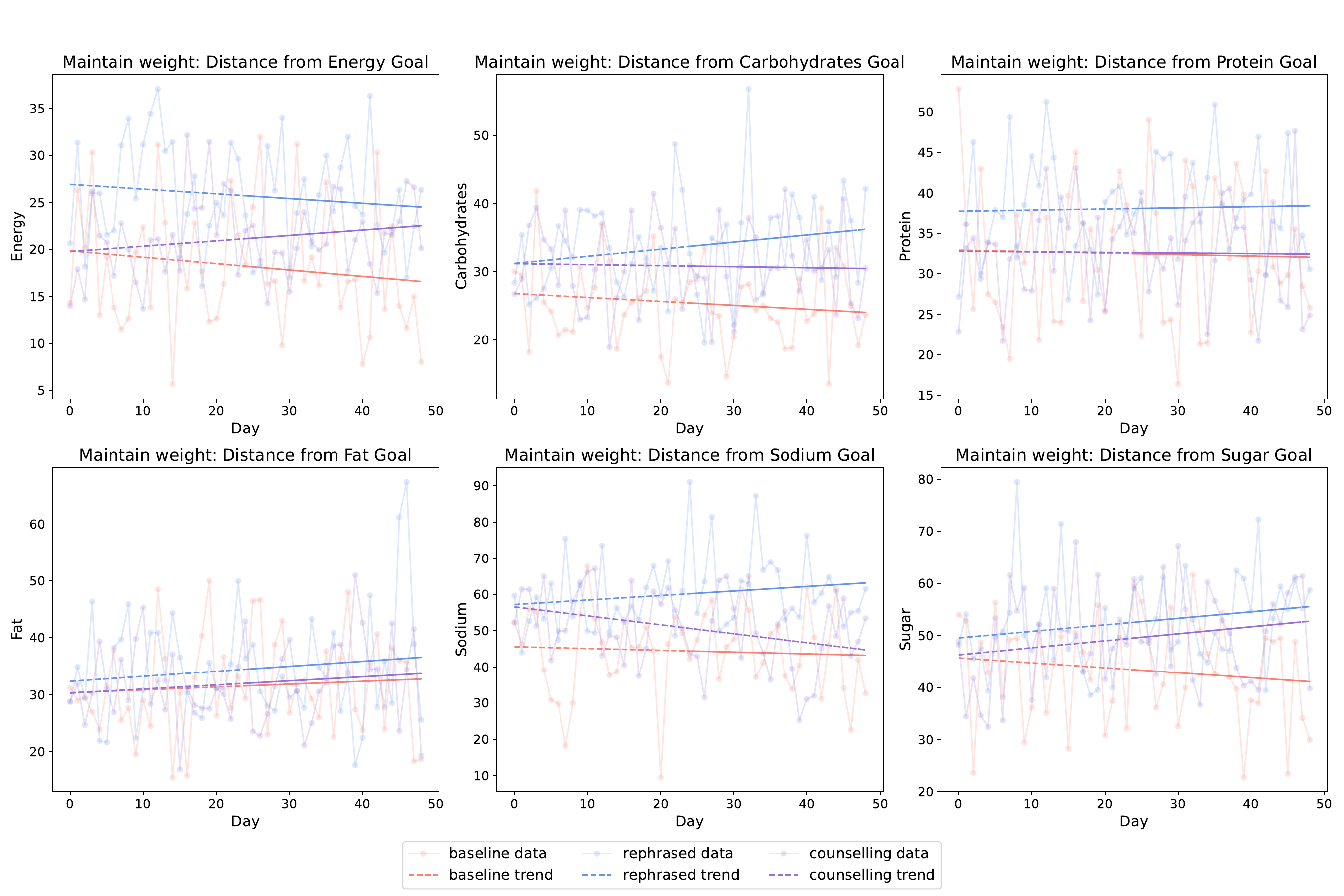}
    \caption{Absolute percentage distance from MyFitnessPal intake goals for participants aiming to maintain their weight.}
    \label{fig:mfp-maintain-distance}
\end{figure*}

\begin{figure*}[ht]
    \includegraphics[width=1\linewidth]{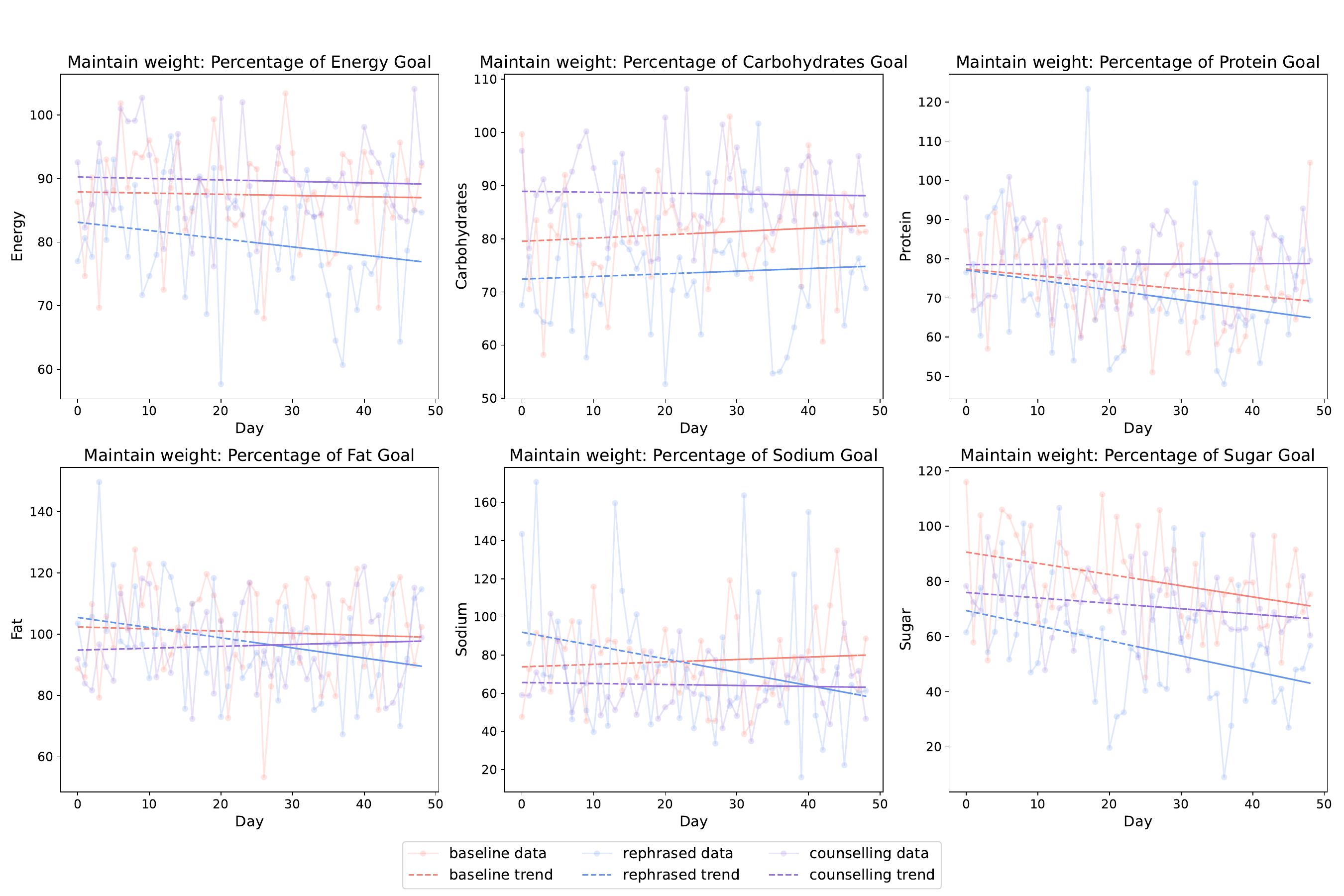}
    \caption{Goal percentage of MyFitnessPal intake goals for participants aiming to maintain their weight.}
    \label{fig:mfp-maintain-percentage}
\end{figure*}

\begin{figure*}[ht]
    \centering
    \includegraphics[width=1\linewidth]{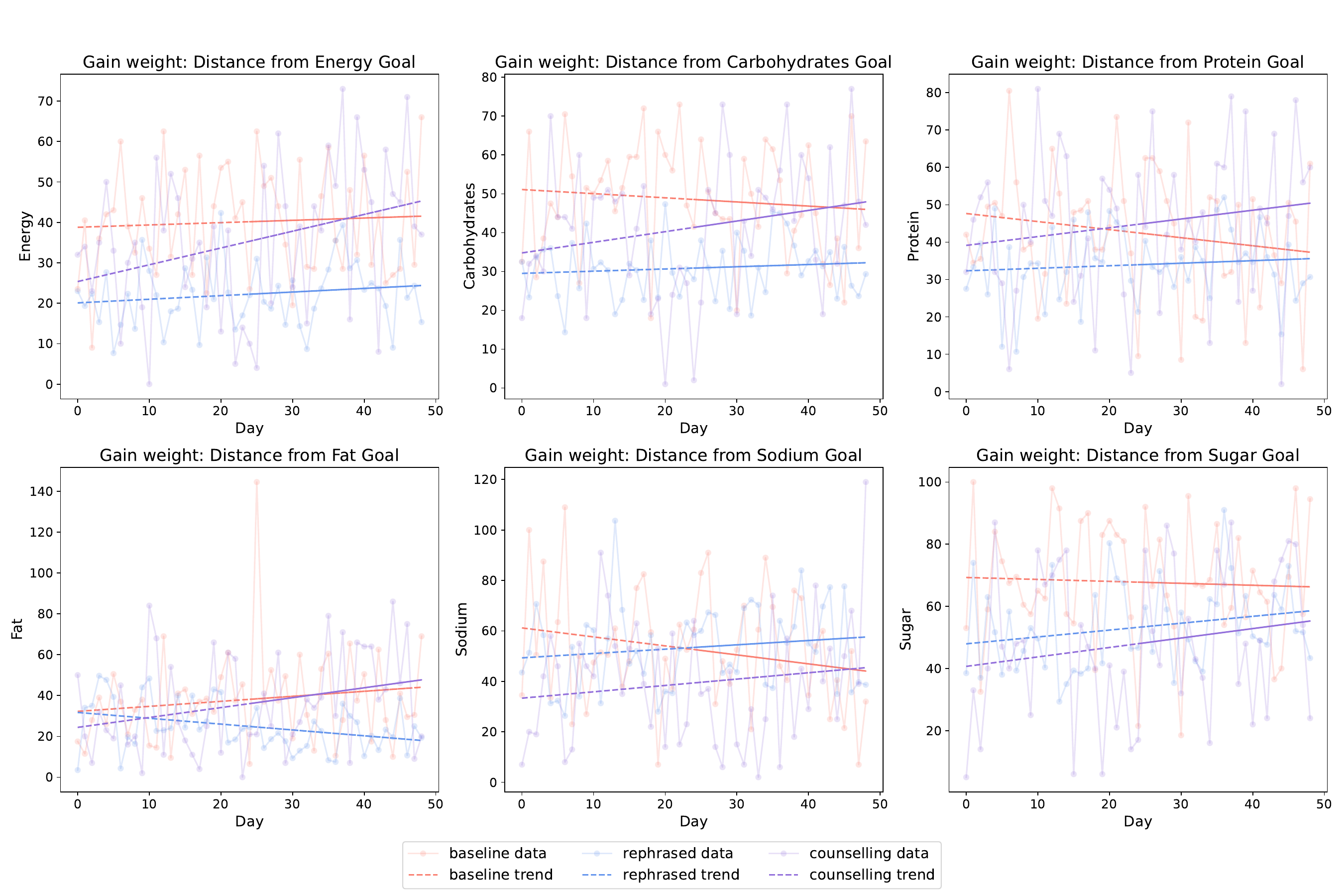}
    \caption{Absolute percentage distance from MyFitnessPal intake goals for participants aiming to gain weight.}
    \label{fig:mfp-gain-distance}
\end{figure*}

\begin{figure*}[ht]
    \includegraphics[width=1\linewidth]{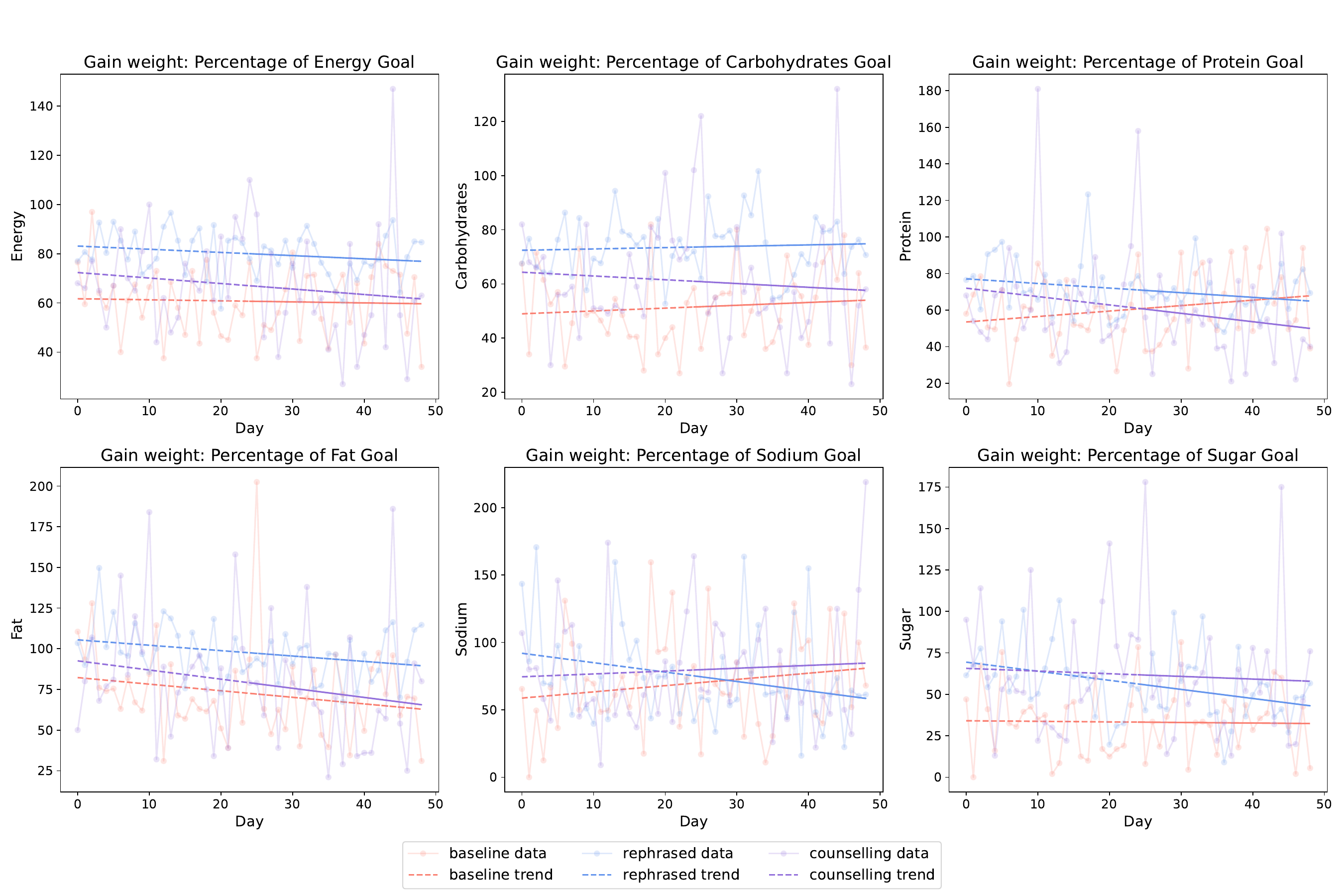}
    \caption{Goal percentage of MyFitnessPal intake goals for participants aiming to gain weight.}
    \label{fig:mfp-gain-percentage}
\end{figure*}

To explore variation in diet outcomes according to the different weight goals, we analysed absolute distance and goal percentage trends among participants aiming to lose, maintain, or gain weight by fitting regression lines to each study group (\Cref{fig:mfp-lose-distance,fig:mfp-lose-percentage,fig:mfp-maintain-distance,fig:mfp-maintain-percentage,fig:mfp-gain-distance,fig:mfp-gain-percentage}). Those seeking to lose weight and receiving nutritional counselling showed greater improvements in goal adherence—especially for energy and protein intake—than other groups. However, participants aiming to gain weight saw poorer adherence across most nutrients in the nutritional counselling group. These subgroup trends suggest that the nutritional counselling feature may have had uneven effects, depending on nutritional goals.

\begin{figure*}
    \centering
    \includegraphics[scale=0.35]{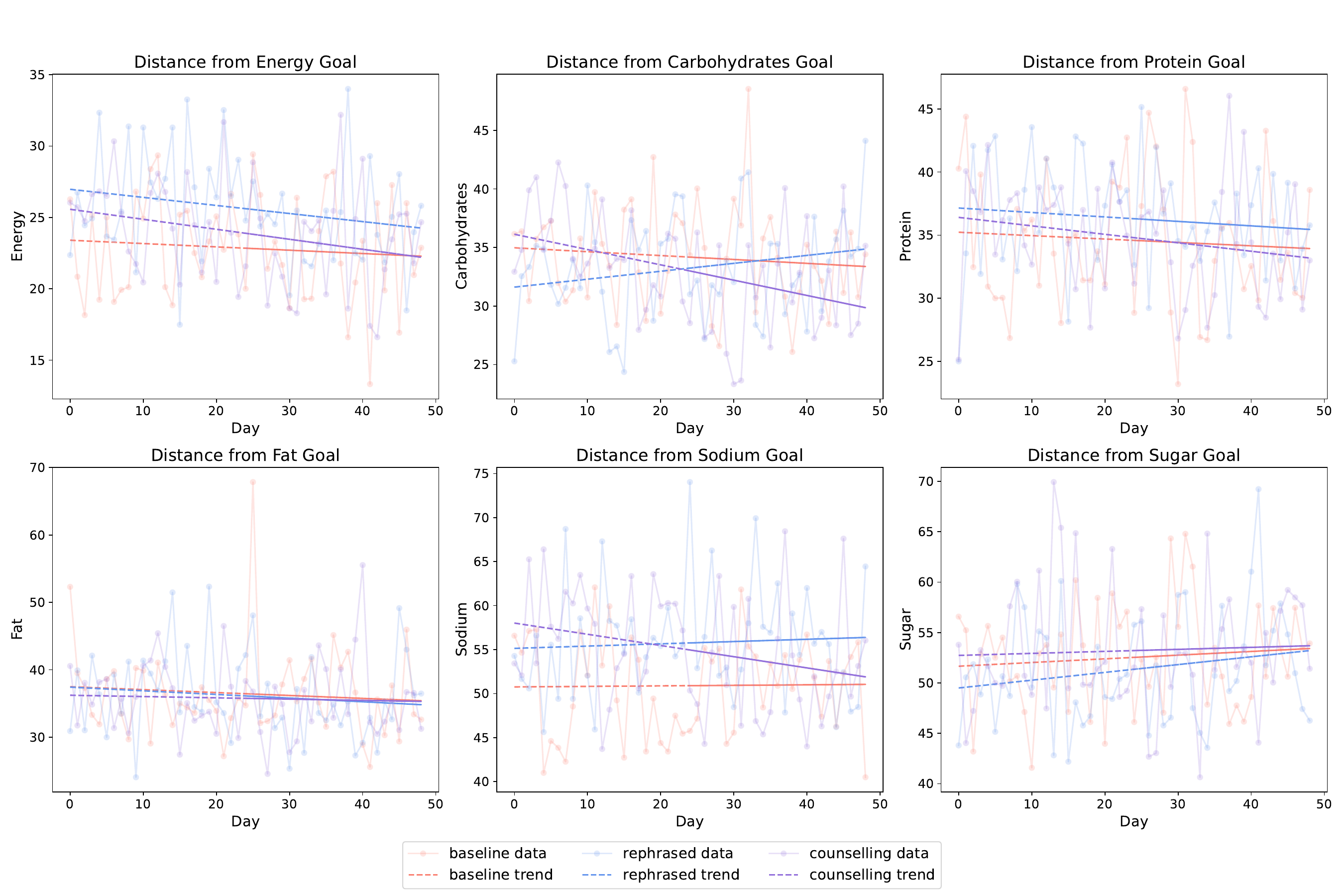}
    \caption{Overall absolute percentage distance from MyFitnessPal intake goals.}
    \label{fig:mfp-overall-distance}
    \centering
    \includegraphics[scale=0.35]{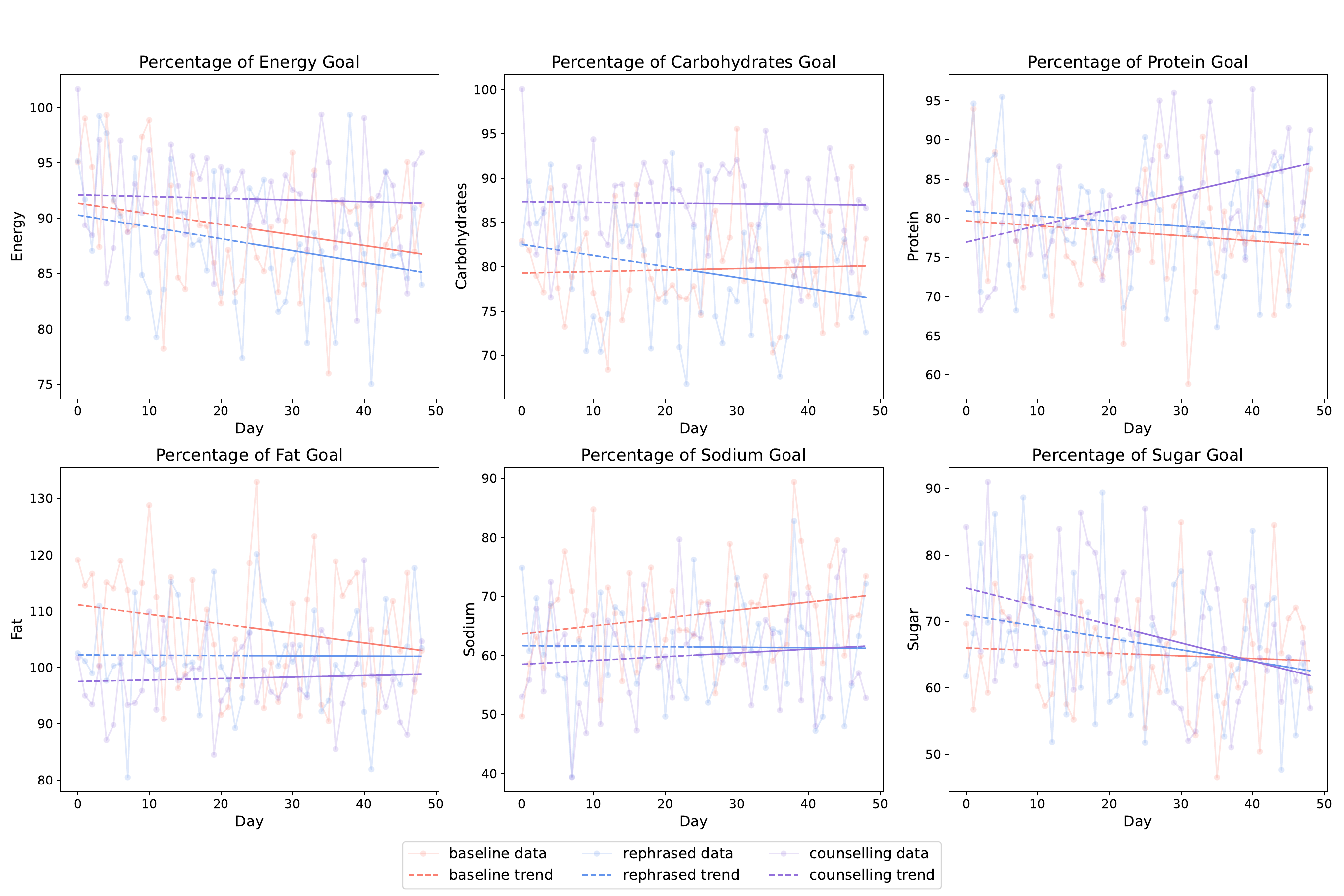}
    \caption{Overall goal percentage of MyFitnessPal intake goals.}
    \label{fig:mfp-overall-percentage}
\end{figure*}

\section{Emotional Well-being Metrics}
\label{app:panas}

\begin{figure*}[ht]
    \centering
    \begin{minipage}{.45\textwidth}
      \centering
      \includegraphics[width=1\linewidth]{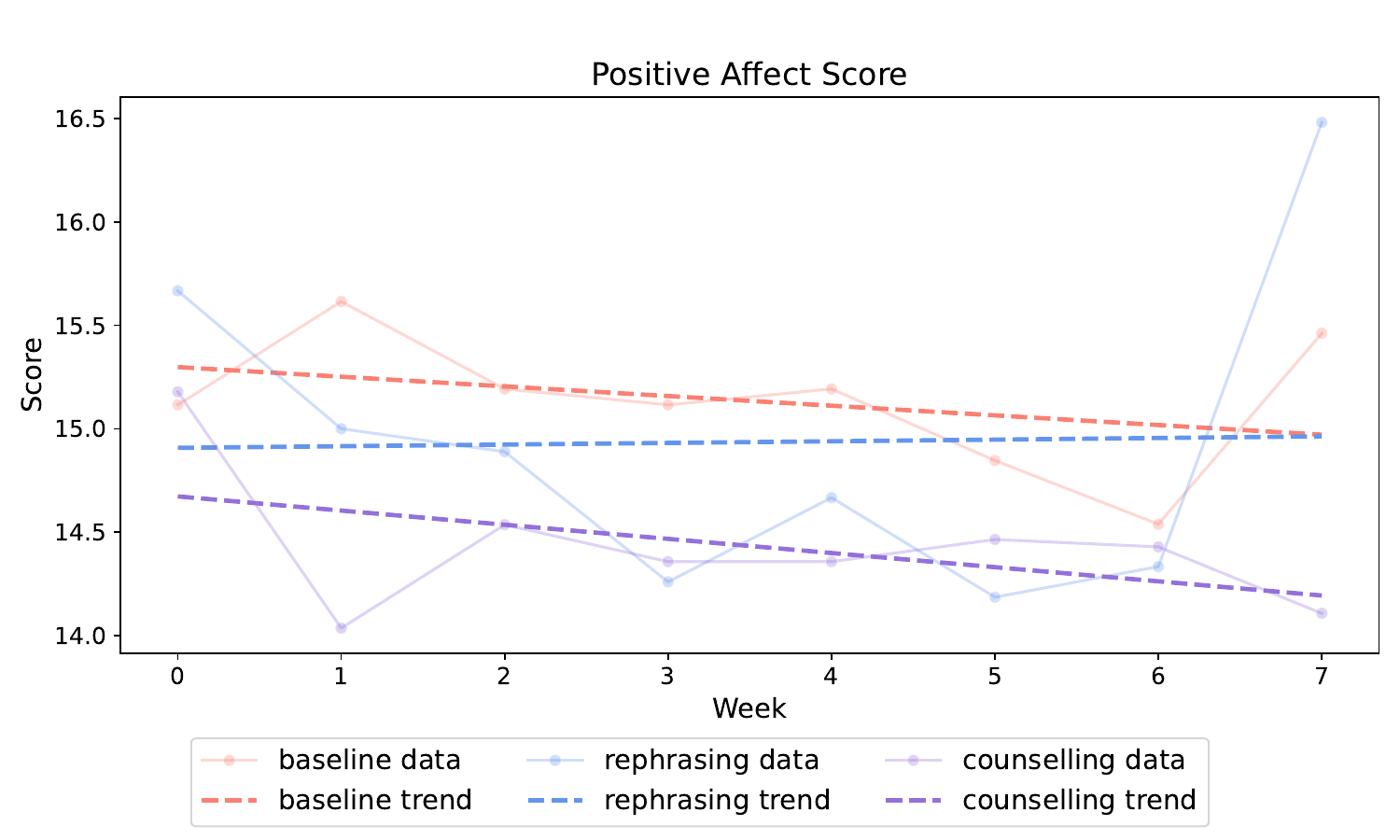}
      \caption{Positive affect score.}
      \label{fig:pos-score}
    \end{minipage}
    \begin{minipage}{.45\textwidth}
      \centering
      \includegraphics[width=1\linewidth]{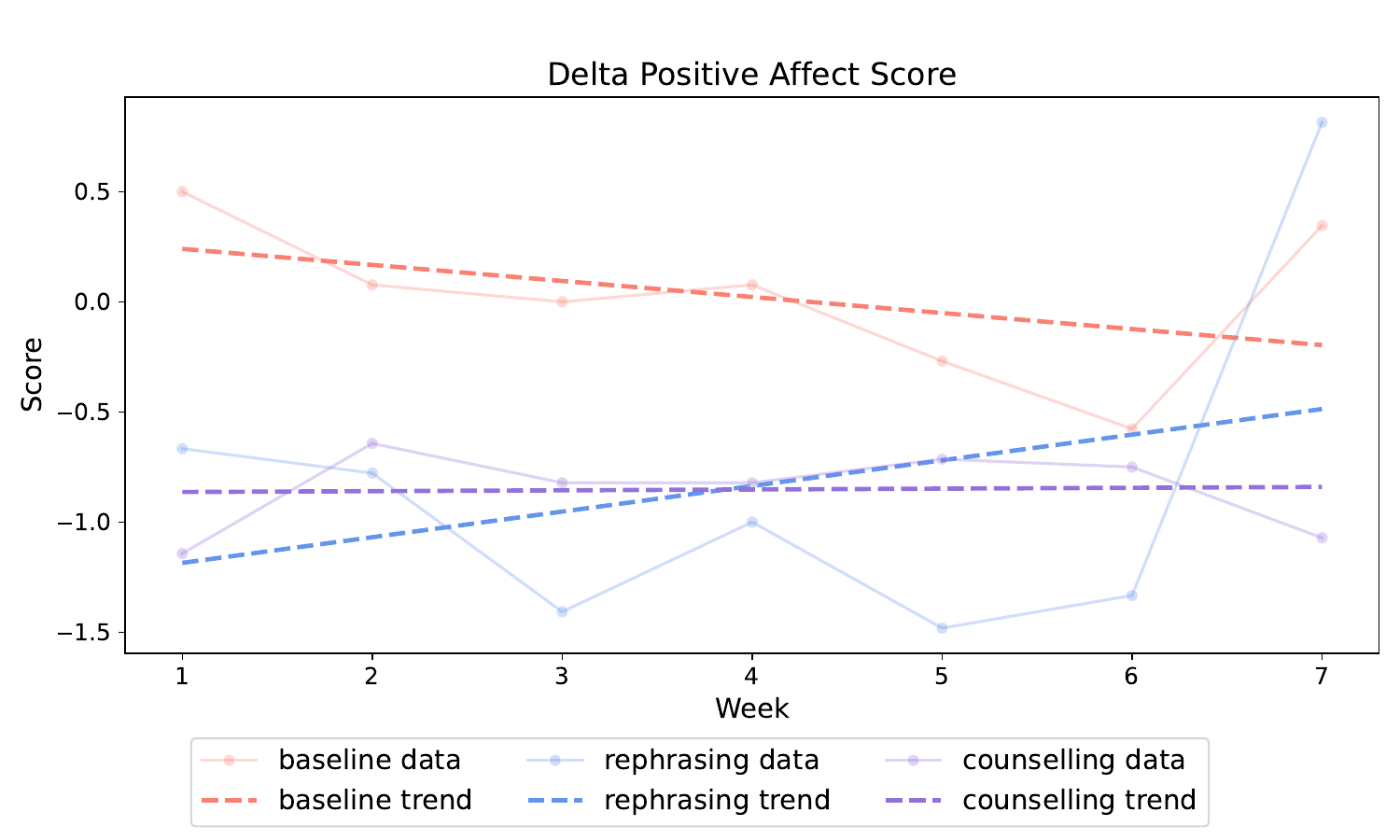}
      \caption{Delta positive affect score.}
      \label{fig:delta-pos-score}
    \end{minipage}
    \includegraphics[width=1\linewidth]{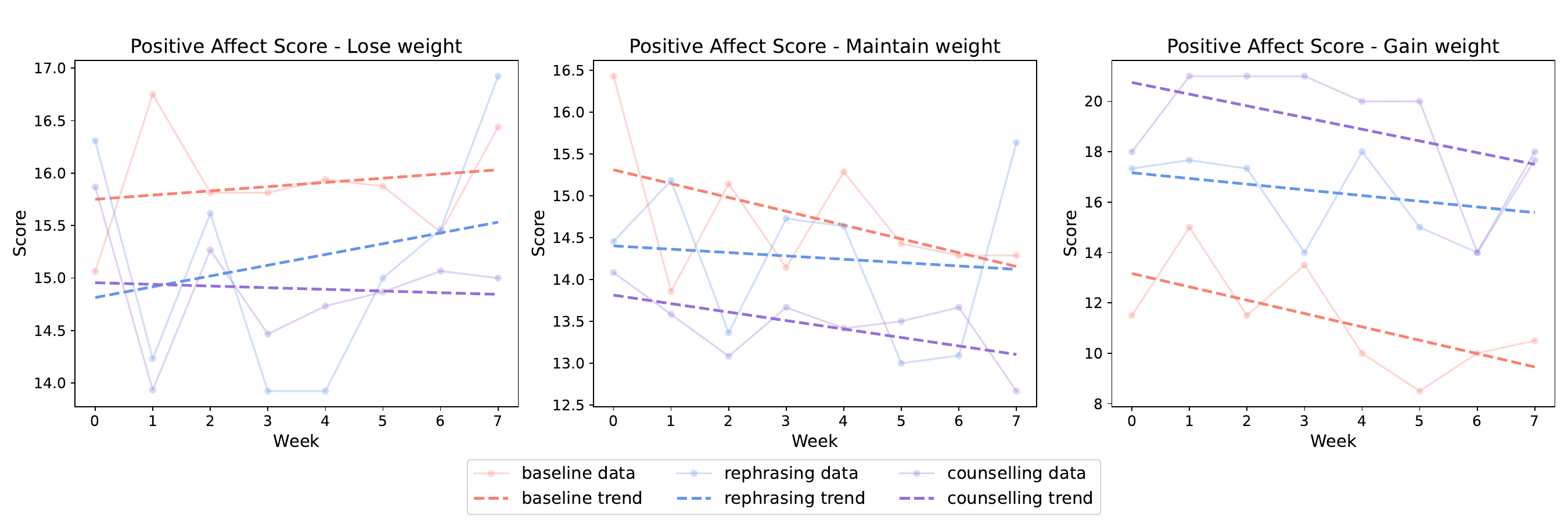}
    \caption{Positive affect score by weight goal.}
    \label{fig:pos-score-by-goal}
    \includegraphics[width=1\linewidth]{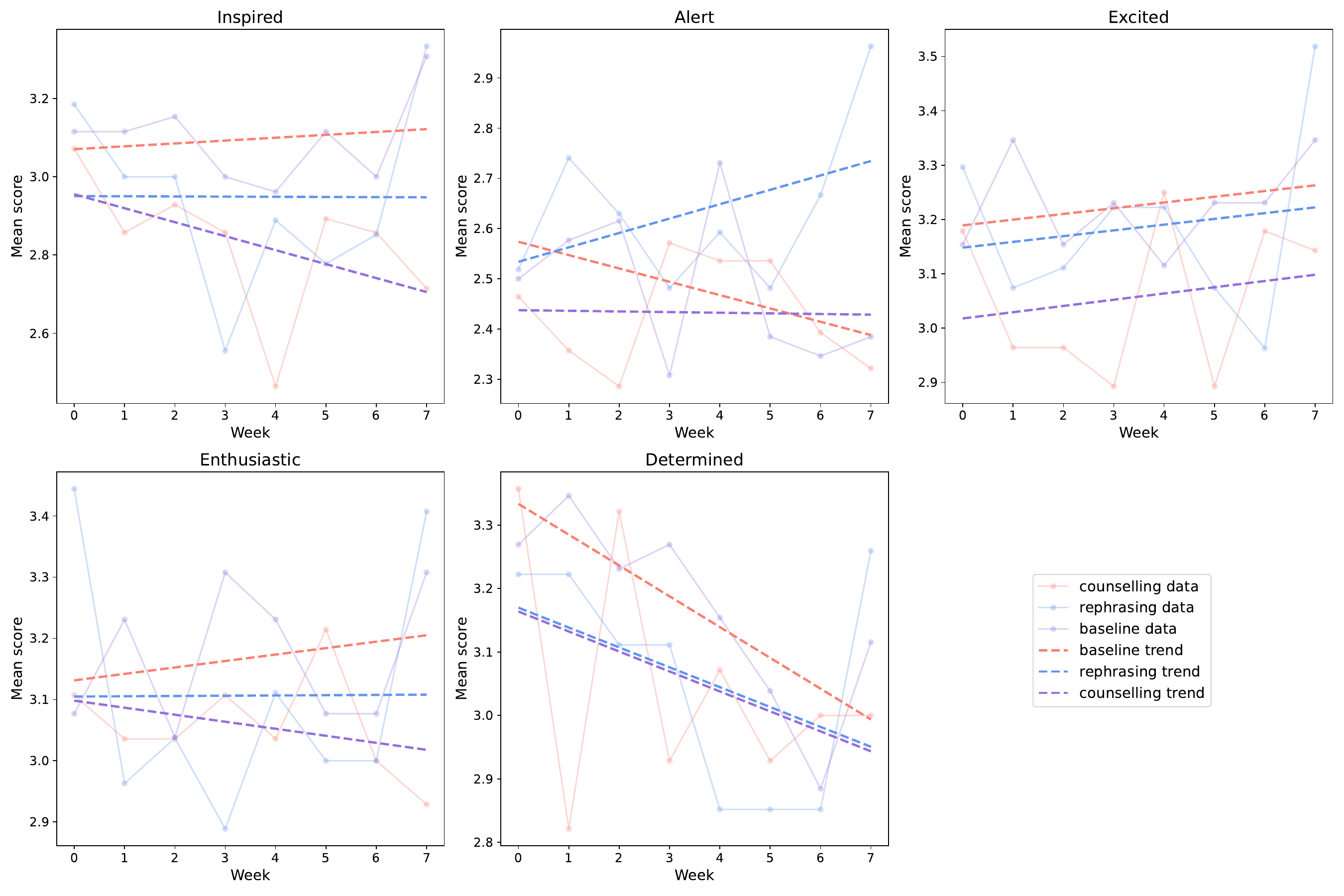}
    \caption{Positive affect score by emotion.}
    \label{fig:pos-score-emotions}
\end{figure*}

In an analysis of the result in \textbf{positive affect} by weight goal subgroups (\Cref{fig:pos-score-by-goal}), we observed a clear divergence in the effectiveness of interventions across groups. Participants with a goal to \textbf{lose weight} demonstrated the most consistent and notable improvements in positive affect scores, particularly in \texttt{REPHRASED}, where scores steadily increased from Week 1 to Week 7. In contrast, those aiming to \textbf{maintain weight} showed more stable and less pronounced changes across all groups. While \texttt{REPHRASED} still outperformed \texttt{BASELINE} and \texttt{FULL}, the magnitude of change for those aiming to maintain their weight was less dramatic compared to those aiming to lose some, suggesting that participants maintaining weight may experience fewer fluctuations in emotional or motivational states due to a less urgent goal. The subgroup of participants looking to \textbf{gain weight}, however, presented more varied outcomes. Positive affect scores fluctuated considerably across weeks, with \texttt{FULL} showing the steepest increases toward the end of the study.

An analysis of \textbf{positive affect} across the specific emotions (\Cref{fig:pos-score-emotions} revealed notable differences between them. While some positive emotions such as ``Alert'' and ``Inspired'' exhibited slight improvements, others, like ``Determined'' and ``Enthusiastic,'' showed either stagnation or a gradual decline over the weeks. This variability suggests that the nutritional counselling intervention in this work may not have effectively addressed the emotional dimensions critical for sustained engagement and motivation.

\begin{figure*}[ht]
    \centering
    \begin{minipage}{.45\textwidth}
      \centering
      \includegraphics[width=1\linewidth]{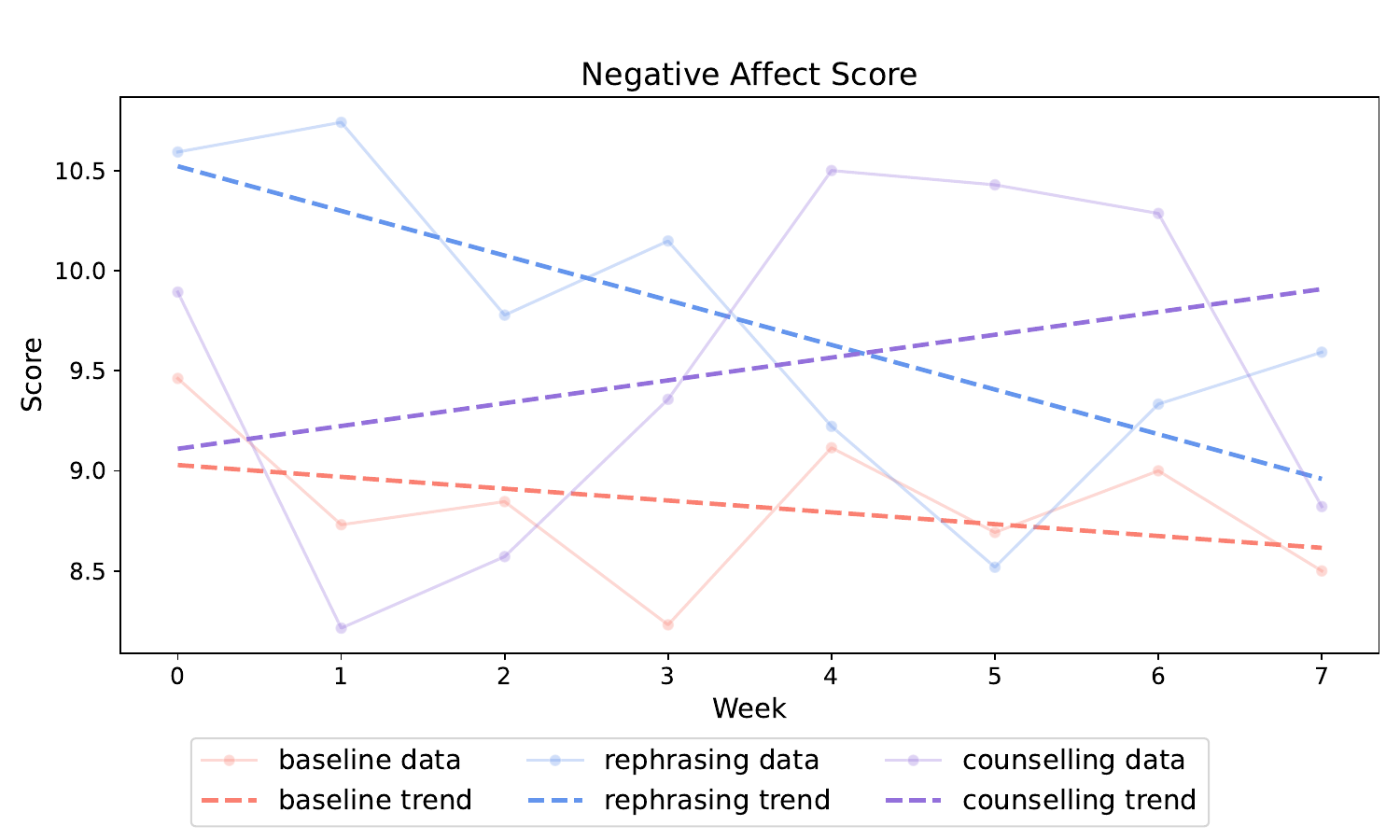}
      \caption{Negative affect score.}
      \label{fig:neg-score}
    \end{minipage}
    \begin{minipage}{.45\textwidth}
      \includegraphics[width=1\linewidth]{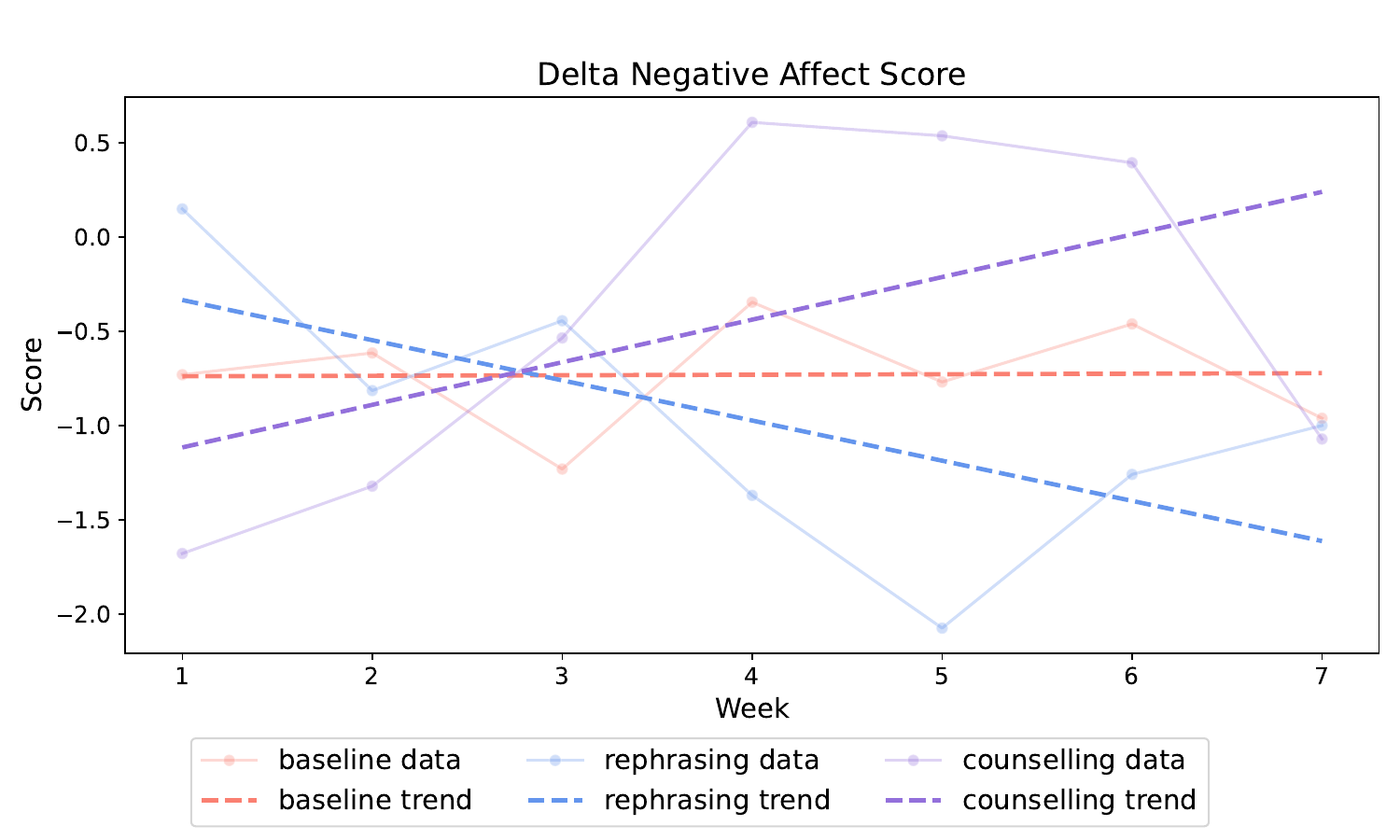}
      \caption{Delta negative affect score.}
      \label{fig:delta-neg-score}
    \end{minipage}
    \includegraphics[width=1\linewidth]{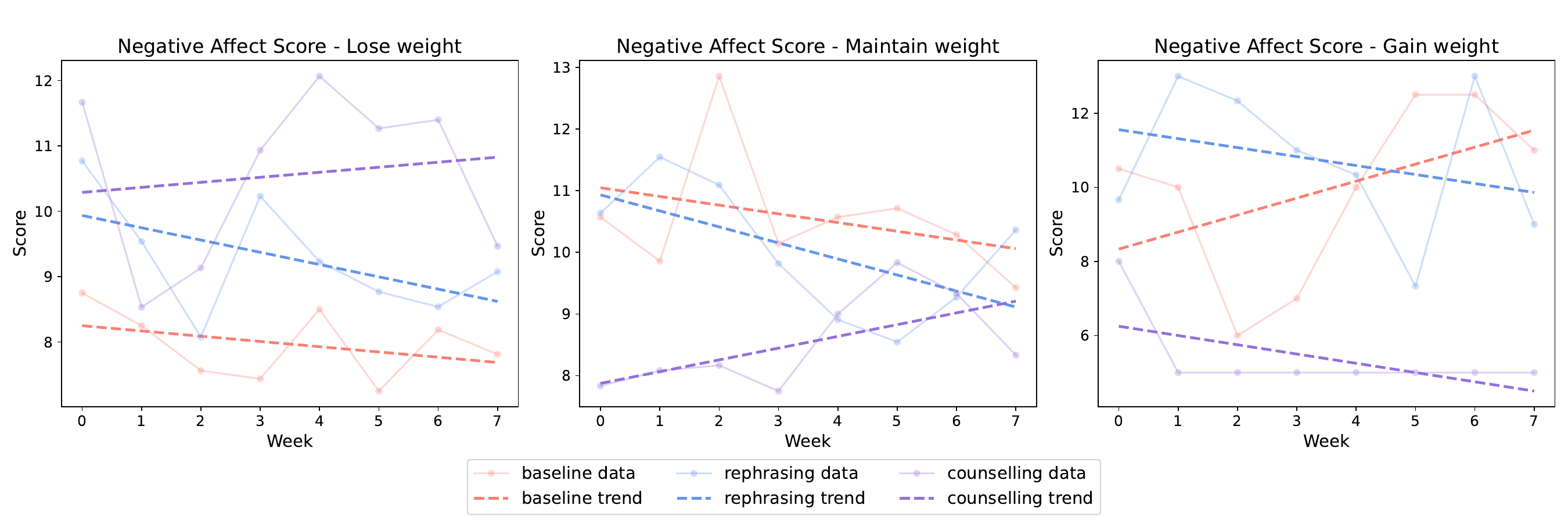}
    \caption{Negative affect score by weight goal.}
    \label{fig:neg-score-by-goal}
    \includegraphics[width=1\linewidth]{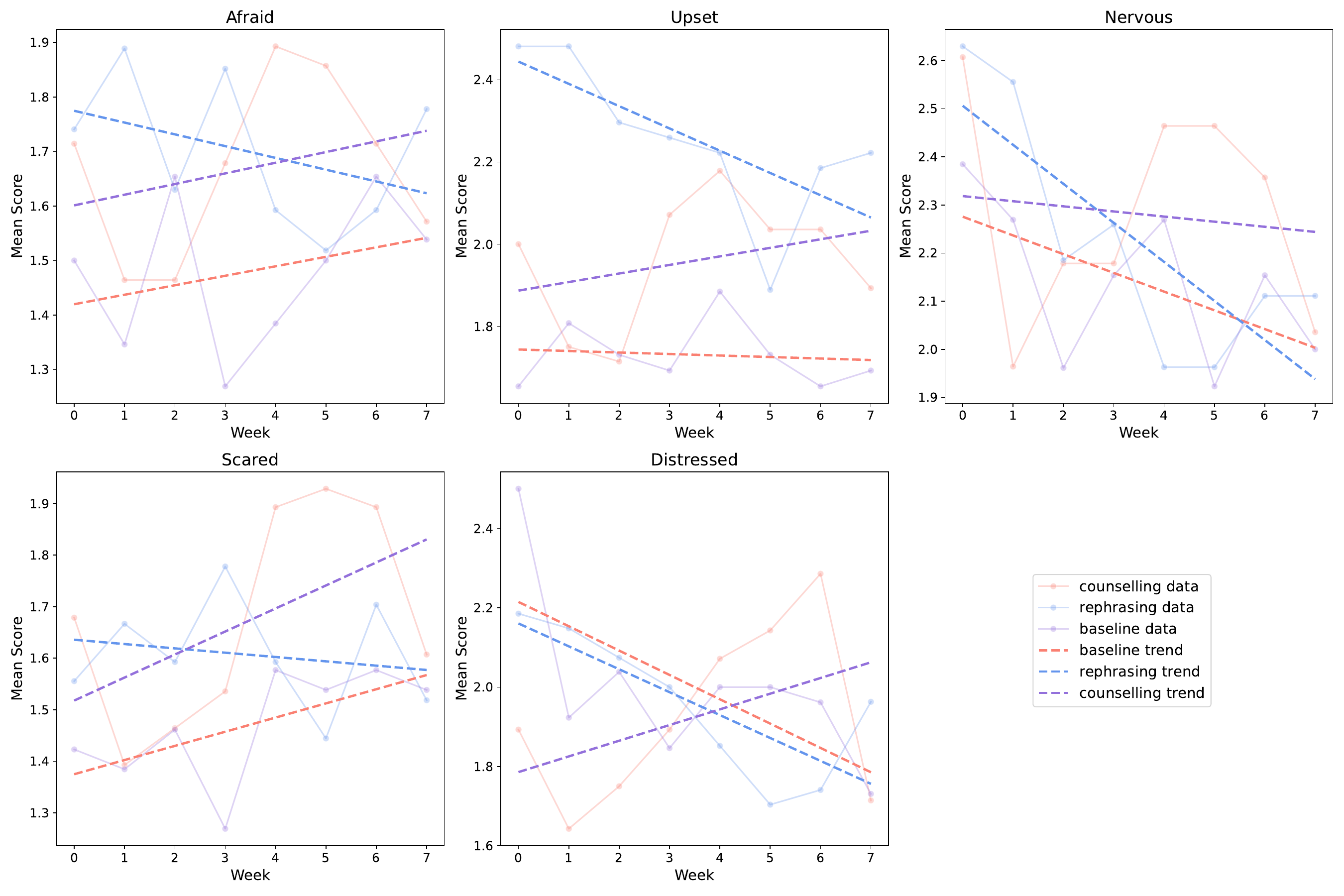}
    \caption{Negative affect score by emotions.}
    \label{fig:neg-score-emotions}
\end{figure*}

Again, we analysed the subgroups with different weight goals for \textbf{negative affect} (\Cref{fig:neg-score-by-goal}). Among participants who aimed to \textbf{lose weight}, those in \texttt{FULL} experienced minimal reductions in negative affect compared to those in \texttt{BASELINE} and \texttt{REPHRASED}, both of which showed significant declines. With participants who wanted to \textbf{maintain weight}, \texttt{FULL} again failed to show improvement, with a relatively flat trend line, while the other two groups showed consistent reductions. For participants aiming to \textbf{gain weight}, \texttt{FULL} exhibited the least improvement, with scores fluctuating without a clear downward trend.

Examining the individual \textbf{negative emotions} (\Cref{fig:neg-score-emotions}) highlights further limitations of nutritional counselling for \texttt{FULL}. The emotions of ``Afraid'', ``Scared'', and ``Upset'' showed little to no improvement in \texttt{FULL}, while the other two groups experienced gradual reductions over the intervention period. Nervousness scores in \texttt{FULL} remained stable or even increased slightly, unlike \texttt{REPHRASED}, which showed a notable decline in this emotion by the end of the study. Despite some fluctuation, there was no significant improvement in distress scores for \texttt{FULL}, in stark contrast to \texttt{REPHRASED}, which showed a steep decline.

\section{User Engagement Metrics}
\label{app:engagement}
\setcounter{figure}{0}
\setcounter{table}{0}

To investigate the effect of the rephrasing of templated responses that was present in both \texttt{REPHRASED} and \texttt{FULL}, we looked at how user engagement differed between these groups and \texttt{BASELINE}. As engagement metrics, we used the interactions (i.e. individual messages from the user), conversations (i.e., a sequence of interactions with responses within five minutes of each other), and days that users interacted with the chatbot, considering the total number and distribution over the trial duration.

\begin{figure*}[ht]
    \centering
    \begin{minipage}{.45\textwidth}
      \centering
      \includegraphics[width=1\linewidth]{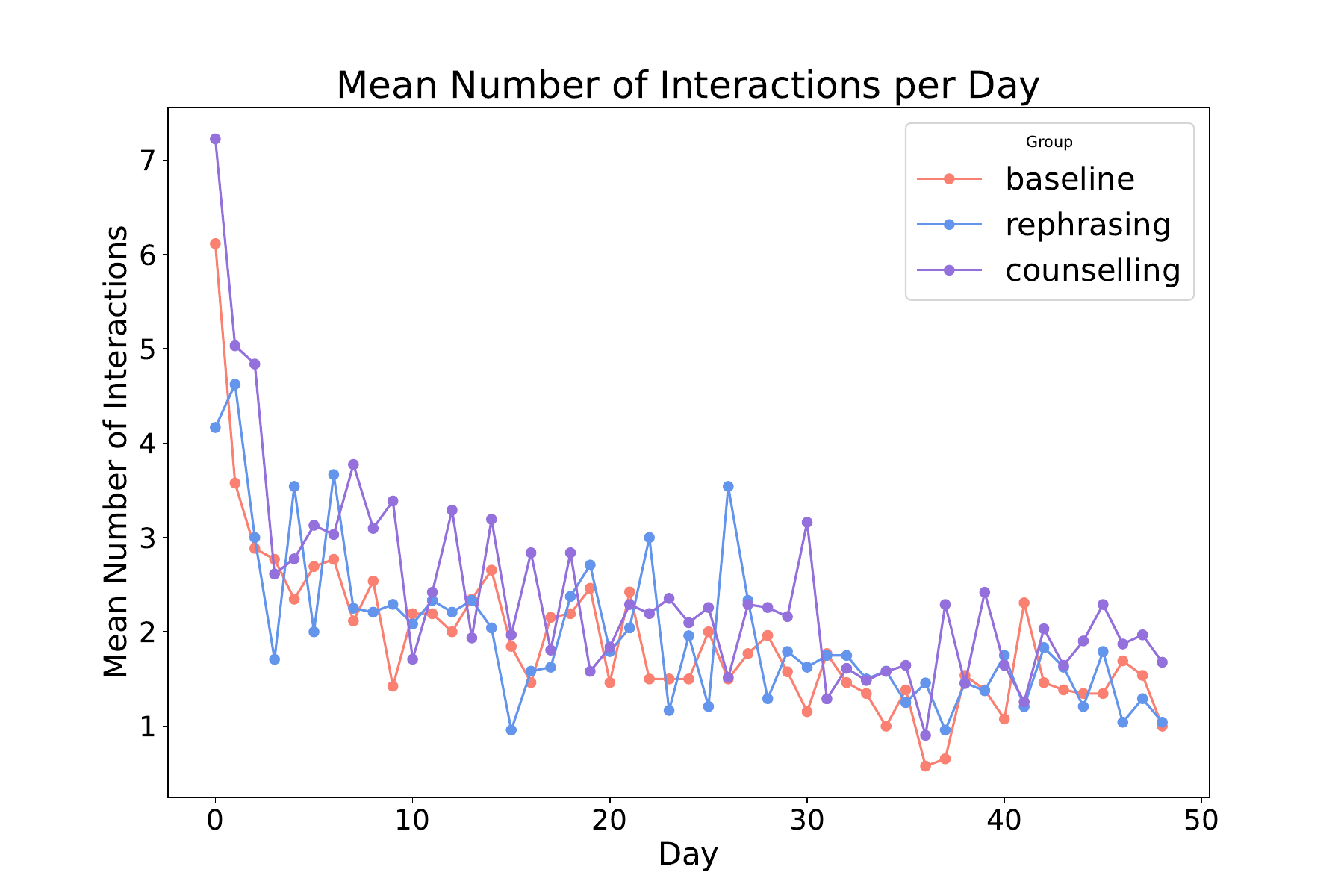}
      \captionof{figure}{Mean number of interactions per day.}
      \label{fig:mean-interactions-per-day}
    \end{minipage}
    \begin{minipage}{.45\textwidth}
      \centering
      \includegraphics[width=1\linewidth]{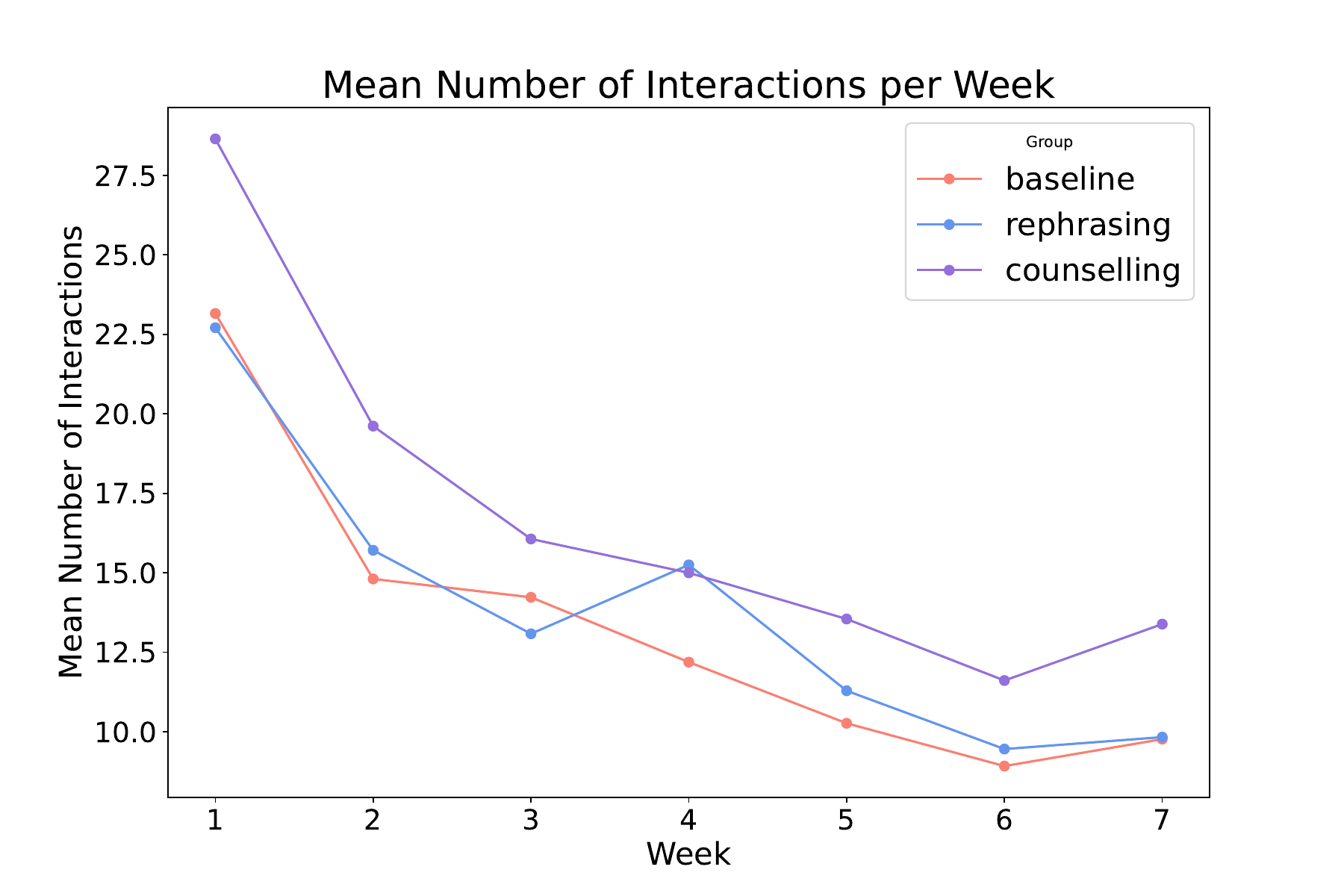}
      \captionof{figure}{Mean number of interactions per week.}
      \label{fig:mean-interactions-per-week}
    \end{minipage}
    \centering
    \begin{minipage}{.45\textwidth}
      \centering
      \includegraphics[width=1\linewidth]{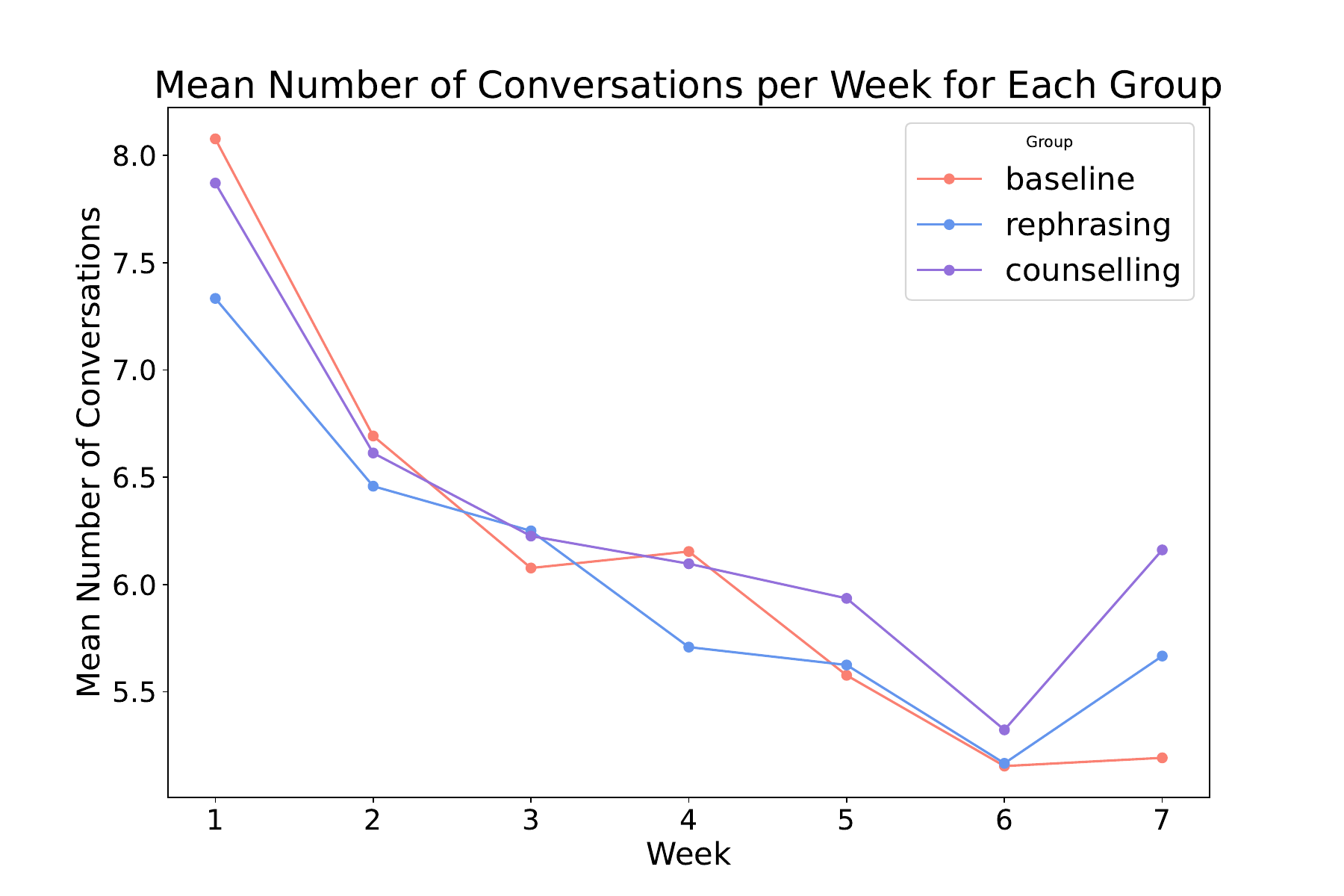}
      \captionof{figure}{Mean number of conversations per week.}
      \label{fig:mean-convos-per-week}
    \end{minipage}
    \begin{minipage}{.45\textwidth}
      \centering
      \includegraphics[width=1\linewidth]{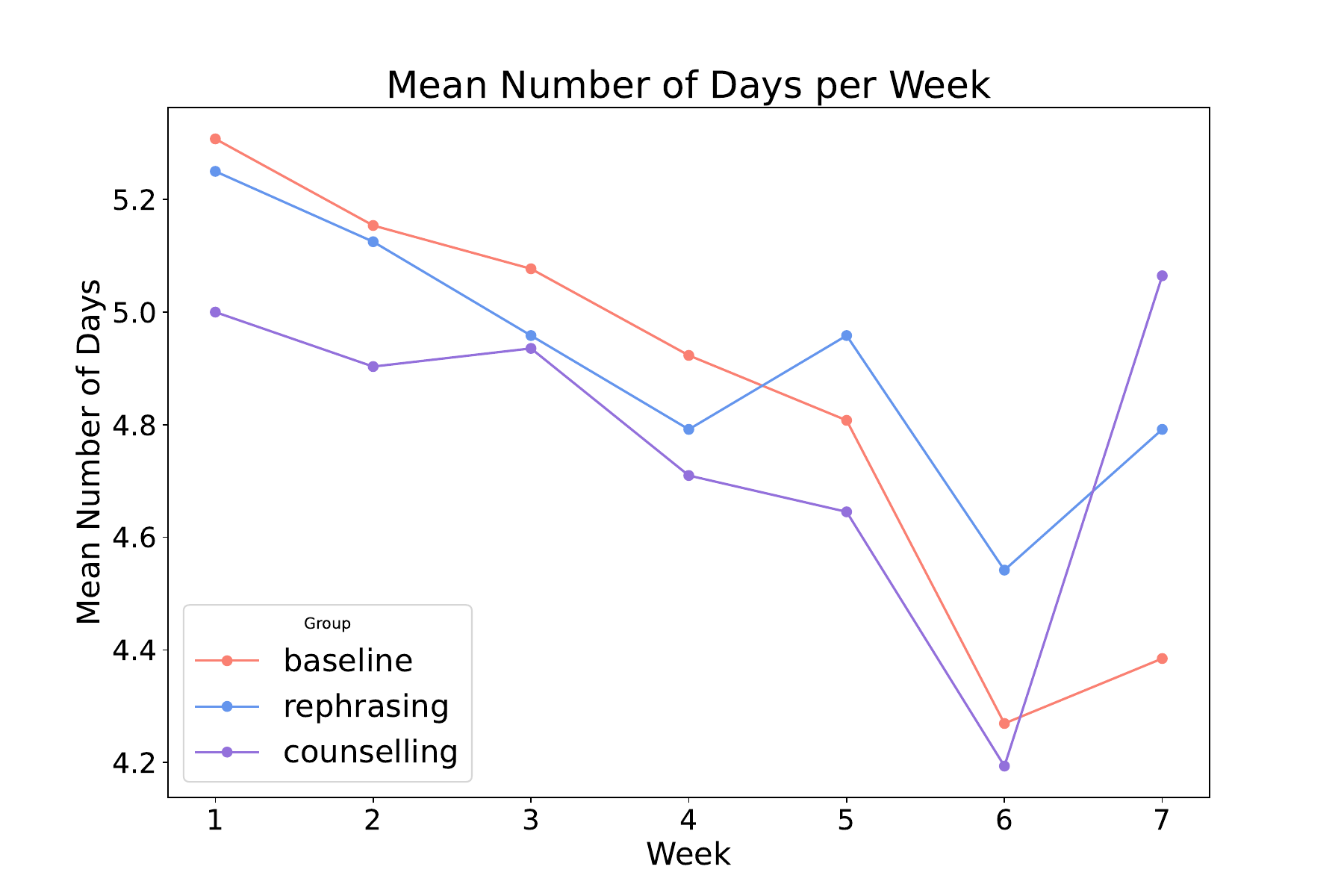}
      \captionof{figure}{Mean number of days users interacted with the chatbot per week.}
      \label{fig:mean-num-days-per-week}
    \end{minipage}
    \centering
    \begin{minipage}{.45\textwidth}
      \centering
      \includegraphics[width=1\linewidth]{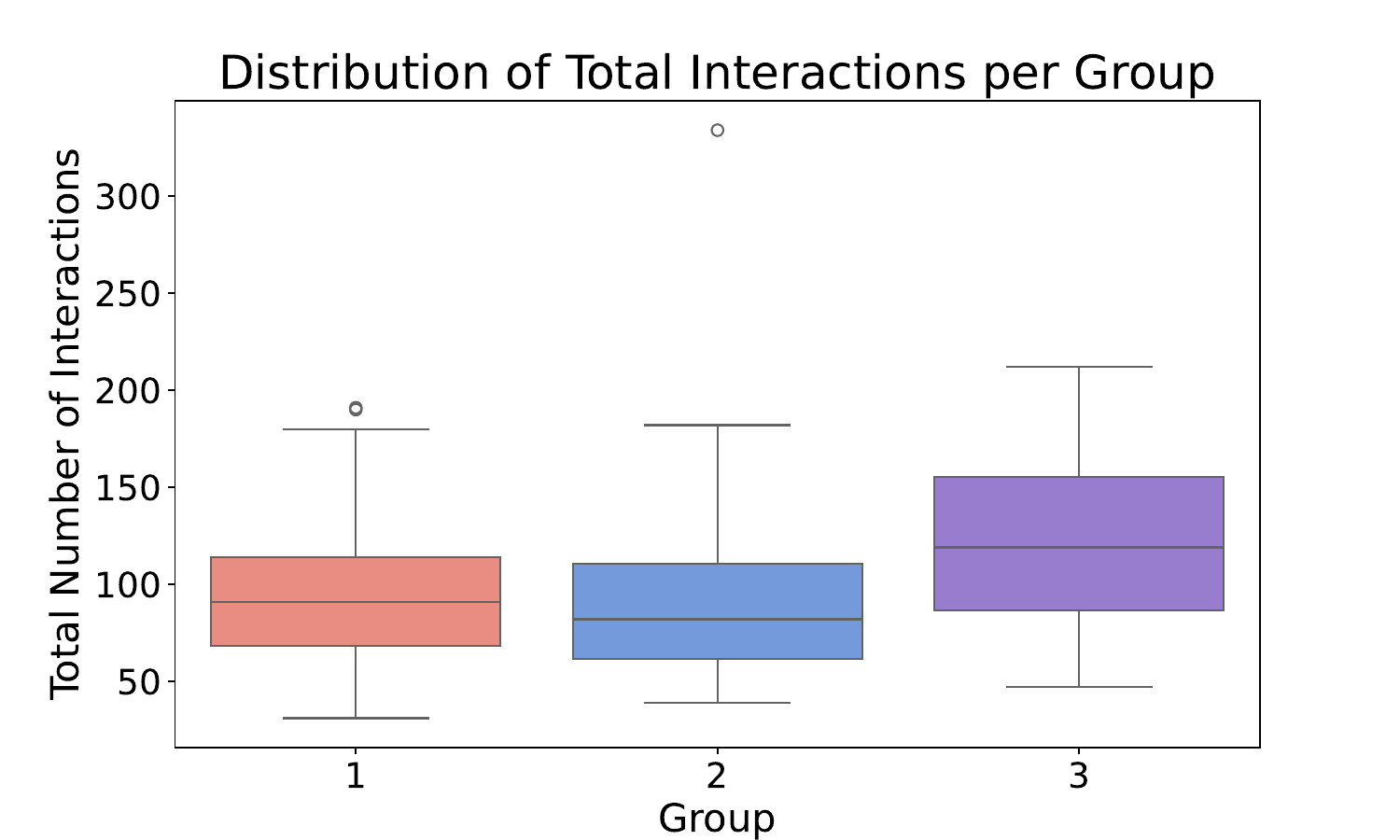}
      \captionof{figure}{Total number of interactions per group.}
      \label{fig:total-interactions-per-group}
    \end{minipage}
    \begin{minipage}{.45\textwidth}
      \centering
      \includegraphics[width=1\linewidth]{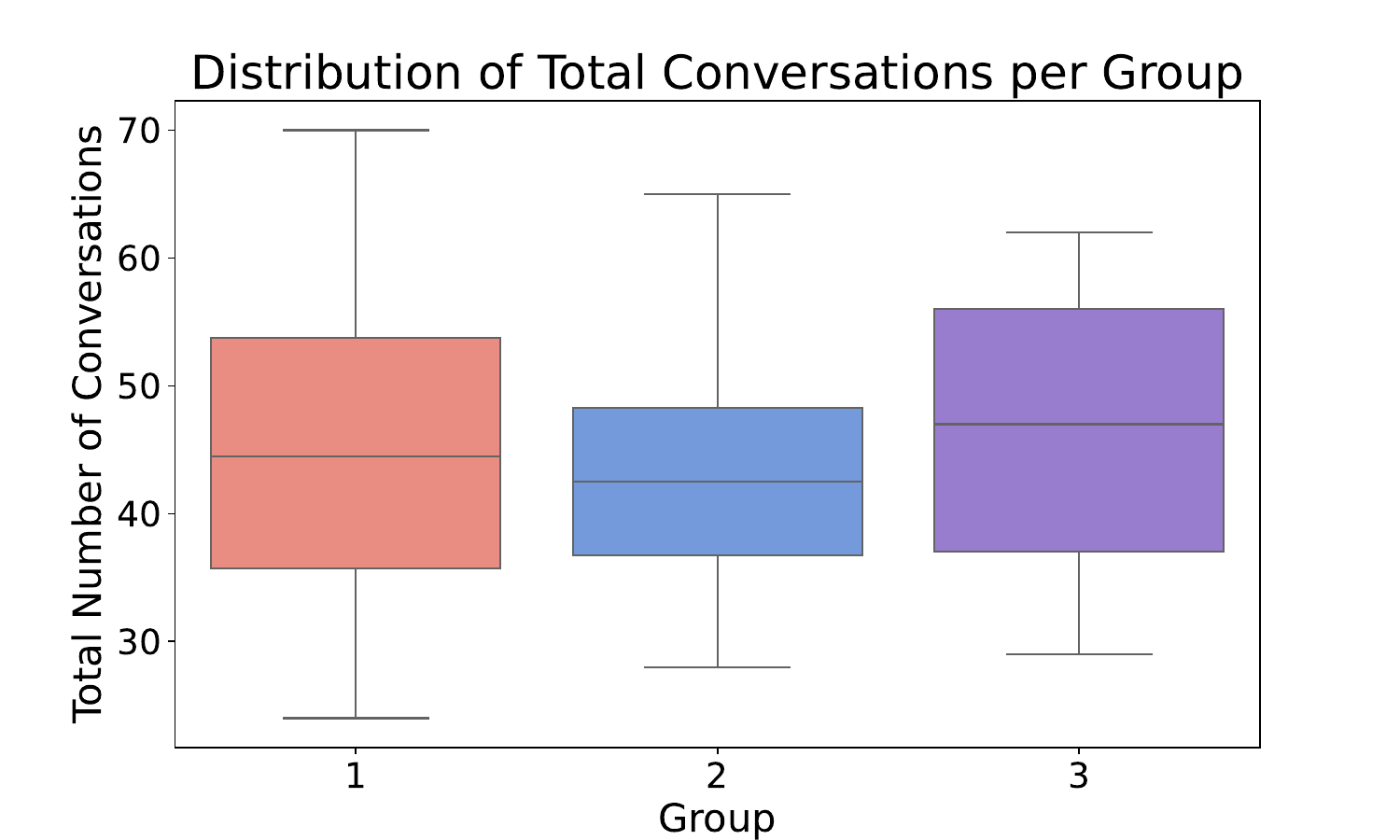}
      \captionof{figure}{Total number of conversations per group.}
      \label{fig:total-conversations-per-group}
    \end{minipage}
\end{figure*}

As shown in \Cref{fig:mean-interactions-per-day,fig:mean-interactions-per-week}, the mean number of interactions per day and week declined consistently over time for all groups, indicating a natural decrease in user engagement as the intervention progressed. However, \texttt{FULL} consistently demonstrated the highest number of interactions overall, likely reflecting the additional ``advice'' feature available exclusively to this group, as well as the prompt to use it every week. In contrast, the rephrased responses alone in \texttt{REPHRASED} did not appear to significantly affect engagement metrics over time compared to \texttt{BASELINE}.

The mean number of conversations per week (\Cref{fig:mean-convos-per-week}) presented more mixed results. While \texttt{FULL} maintained the highest number of conversations, as highlighted by the total number of conversations in \Cref{fig:total-conversations-per-group}, the differences were not as pronounced as the total number of interactions (\Cref{fig:total-interactions-per-group}). The lack of notable differences between \texttt{BASELINE} and \texttt{REPHRASED} suggests that the increase in overall conversations observed in \texttt{FULL} was also driven primarily by the additional ``advice'' feature, rather than the rephrased responses shared by \texttt{REPHRASED} and \texttt{FULL}.  Furthermore, \texttt{FULL} did not engage with the chatbot on more days per week than the other groups (\Cref{fig:mean-num-days-per-week}), so there was no clear evidence that the rephrased responses had any consistent impact on this metric.

These findings suggest that while the ``advice'' feature and the weekly prompt in \texttt{FULL} contributed to overall engagement, rephrased responses did not significantly influence the number of conversations or the frequency of chatbot use.

\end{document}